\documentclass[12pt]{article}
\usepackage{amsmath,amssymb}
\newcommand{\eqdef}{\stackrel{\text{def}}{=}}
\newcommand{\n}{\nonumber\\}
\newcommand{\bm}{\boldsymbol}
\newcommand{\ignore}[1]{}

\newcommand{\Romannumeral}[1]{\uppercase\expandafter{\romannumeral#1}}

\allowdisplaybreaks[4]

\begin{document}

\baselineskip=20pt

%%%%%%%%%%%%%%%%%%%%%%%%%%%%%%%%%%%%%%%%%%%%%%%%%%%%%%%%%%%%
%                                                          %
%  Title page                                              %
%                                                          %
%%%%%%%%%%%%%%%%%%%%%%%%%%%%%%%%%%%%%%%%%%%%%%%%%%%%%%%%%%%%
\newfont{\elevenmib}{cmmib10 scaled\magstep1}
\newcommand{\preprint}{
    \begin{flushleft}
     \elevenmib Yukawa\, Institute\, Kyoto\\
   \end{flushleft}\vspace{-1.3cm}
   \begin{flushright}\normalsize \sf
     DPSU-11-2\\
     YITP-11-18\\
%     {\tt arXiv:1102.nnnn[math-ph]}\\
%     February 2010
   \end{flushright}}
\newcommand{\Title}[1]{{\baselineskip=26pt
   \begin{center} \Large \bf #1 \\ \ \\ \end{center}}}
\newcommand{\Author}{\begin{center}
   \large \bf Satoru Odake${}^a$ and Ryu Sasaki${}^b$ \end{center}}
\newcommand{\Address}{\begin{center}
     $^a$ Department of Physics, Shinshu University,\\
     Matsumoto 390-8621, Japan\\
     ${}^b$ Yukawa Institute for Theoretical Physics,\\
     Kyoto University, Kyoto 606-8502, Japan
   \end{center}}
\newcommand{\Accepted}[1]{\begin{center}
   {\large \sf #1}\\ \vspace{1mm}{\small \sf Accepted for Publication}
   \end{center}}

\preprint
\thispagestyle{empty}

\Title{The Exceptional ($X_{\ell}$) ($q$)-Racah Polynomials}
\Author

\Address
\vspace{1cm}

\begin{abstract}
The exceptional Racah and $q$-Racah polynomials are constructed.
Together with the exceptional Laguerre, Jacobi, Wilson and Askey-Wilson
polynomials discovered by the present authors in 2009, they exhaust the
generic exceptional orthogonal polynomials of a single variable.
\end{abstract}

%%%%%%%%%%%%%%%%%%%%%%%%%%%%%%%%%%%%%%%%%%%%%%%%%%%%%%%%%%%%%%%
%                                                             %
%  1. Introduction                                            %
%                                                             %
%%%%%%%%%%%%%%%%%%%%%%%%%%%%%%%%%%%%%%%%%%%%%%%%%%%%%%%%%%%%%%%
\section{Introduction}
\label{intro}

The exceptional ($X_\ell$) ($q$)-Racah polynomials and related exceptional
orthogonal polynomials are constructed as the main part of the eigenfunctions
of the shape invariant and exactly solvable discrete quantum mechanics with
real shifts \cite{os12}, which are deformations of those governing the
corresponding orthogonal polynomials, {\em i.e.\/}\ the ($q$)-Racah
polynomials, etc. \cite{nikiforov,askey,ismail,koeswart}.
The method of deformations is essentially the same as that for the ($X_\ell$) 
Wilson and Askey-Wilson polynomials derived by the present authors in
2009 \cite{os17}. Namely, the potential functions of the original
Hamiltonians are multiplicatively deformed in terms of a degree $\ell$
eigenpolynomial with twisted parameters. The exceptional ($q$)-Racah
polynomials and the exceptional Wilson and Askey-Wilson polynomials share
many properties.
One pronounced difference is that there are only finitely many exceptional
($q$)-Racah polynomials in contrast with the infinitely many types of
the exceptional Wilson and Askey-Wilson polynomials.
For example, starting from the ($q$)-Racah polynomials of the highest
degree $N$, there exist $N-1$ different types of the exceptional ($q$)-Racah
polynomials, for which the highest degree is always $N$.
On the other hand, there are infinitely many different types of the
exceptional little $q$-Jacobi polynomials, since the degrees of the
original little $q$-Jacobi polynomials are not bounded.
These exceptional ($X_\ell$) polynomials are {\em exceptional\/} in the
sense that they form a complete set of orthogonal polynomials in spite of
the fact that the lowest member of the polynomials has degree $\ell$
($\ge1$) instead of a constant. Thus they do not satisfy the three term
recurrence relations.

Historically the $X_1$ Laguerre and Jacobi polynomials were discovered
by G\'omez-Ullate et al \cite{gomez} in 2008 within the framework of the
Sturm-Liouville theory. Soon they were rederived as the main part of the
eigenfunctions of shape invariant quantum mechanical Hamiltonians by Quesne
and collaborators \cite{quesne}. In 2009 the present authors derived the
infinitely many $X_\ell$ Laguerre and Jacobi polynomials by deforming the
Hamiltonian systems of the radial oscillator and the P\"oschl-Teller
potential in terms of the eigenpolynomials of degree $\ell$
\cite{os16,os18,os19}.
The examples of G\'omez-Ullate et al and Quesne et al are the first
members of the infinitely many exceptional polynomials. For the recent
developments of the exceptional orthogonal polynomials, see
\cite{hos,duttaroy,stz,os20,os21}.
It is worth remarking that the general knowledge of the solution spaces of
exactly solvable (discrete) quantum mechanical systems governed by Crum's
theorem \cite{crum} and its modifications \cite{adler,os15,gos,os22} has
been very helpful for the discovery of various exceptional orthogonal
polynomials.

The orthogonal polynomials of a discrete variable \cite{nikiforov} have
played important roles in many disciplines of physics and mathematics
\cite{nikiforov,askey,ismail}. See \cite{albanese} for recent applications.
Let us comment on the birth and death processes, the typical examples of
Markov chains, which could be considered as a discrete version of the
Fokker-Planck equations \cite{risken}. As shown in \cite{bdproc}, the
explicit examples of 18 orthogonal polynomials in \cite{os12}, the
($q$)-Racah, ($q$)-(dual)-Hahn etc, provide {\em exactly solvable birth
and death processes\/} \cite{ismail,bdp}. That is, for the given birth
and death rates $\{B(x),D(x)\}$ which define the Hamiltonian \eqref{genham},
the corresponding transition probabilities are given explicitly, not in
a general spectral representation form of Karlin-McGregor \cite{karlin}.
The exceptional versions presented here also provides ample examples of
exactly solvable birth and death processes.

The present paper is organised as follows. In section two, the basic
principles of the discrete quantum mechanics with real shifts are briefly
reviewed with an emphasis on the shape invariance. The details of the Racah
and $q$-Racah polynomials are recapitulated in section three.
The exceptional Racah and $q$-Racah polynomials are introduced in section
four. The intertwining relations connecting the original ($q$)-Racah and
the exceptional ($q$)-Racah polynomials are explored in section five.
These two sections are the main part of this paper.
Several exceptional orthogonal polynomials, the dual ($q$)-Hahn and the
little $q$-Jacobi polynomials are derived from the exceptional ($q$)-Racah
polynomials in section six through certain limiting processes. The final
section is for a summary and comments. 

%%%%%%%%%%%%%%%%%%%%%%%%%%%%%%%%%%%%%%%%%%%%%%%%%%%%%%%%%%%%%%%
%                                                             %
%  2. General Setting: shape invariance                       %
%                                                             %
%%%%%%%%%%%%%%%%%%%%%%%%%%%%%%%%%%%%%%%%%%%%%%%%%%%%%%%%%%%%%%%
\section{General Setting: shape invariance}
\label{sec:dQM}
\setcounter{equation}{0}

Let us recapitulate the essence of  the discrete quantum mechanics with
real shifts developed in \cite{os12}.
The Hamiltonian $\mathcal{H}=(\mathcal{H}_{x,y})$ is a tridiagonal real
symmetric (Jacobi) matrix and its rows and columns are indexed by non-negative
integers $x$ and $y$, $x,y=0,1,\ldots,x_{\text{max}}$, either finite
($x_{\text{max}}=N$) or infinite ($x_{\text{max}}=\infty$).
The Hamiltonian $\mathcal{H}$ has a form
\begin{align}
  &\mathcal{H}\eqdef
  -\sqrt{B(x)}\,e^{\partial}\sqrt{D(x)}-\sqrt{D(x)}\,e^{-\partial}\sqrt{B(x)}
  +B(x)+D(x),
  \label{genham}\\
  &\mathcal{H}_{x,y}=
  -\sqrt{B(x)D(x+1)}\,\delta_{x+1,y}-\sqrt{B(x-1)D(x)}\,\delta_{x-1,y}
  +\bigl(B(x)+D(x)\bigr)\delta_{x,y},
\end{align}
in which the two functions $B(x)$ and $D(x)$ are real and {\em positive}
but vanish at the boundary:
\begin{align}
  B(x)>0,\quad D(x)>0,\quad D(0)=0\ ;\quad
  B(x_{\text{max}})=0\ \ \text{for finite case}.
  \label{BDcondition}
\end{align}
The Schr\"{o}dinger equation is the eigenvalue problem for a hermitian
matrix $\mathcal{H}$ ($n_{\text{max}}=N$ or $\infty$),
\begin{equation}
  \mathcal{H}\phi_n(x)=\mathcal{E}_n\phi_n(x)\quad
  (n=0,1,\ldots,n_{\text{max}}),\quad
  0=\mathcal{E}_0<\mathcal{E}_1<\mathcal{E}_2<\cdots.
  \label{Hspec}
\end{equation}
The Hamiltonian \eqref{genham} can be expressed in a factorised form:
\begin{align}
  &\mathcal{H}=\mathcal{A}^{\dagger}\mathcal{A},\qquad
  \mathcal{A}=(\mathcal{A}_{x,y}),
  \ \ \mathcal{A}^{\dagger}=((\mathcal{A}^{\dagger})_{x,y})
  =(\mathcal{A}_{y,x}),
  \ \ (x,y=0,1,\ldots,x_{\text{max}}),
  \label{factor}\\
  &\mathcal{A}\eqdef\sqrt{B(x)}-e^{\partial}\sqrt{D(x)},\quad
  \mathcal{A}^{\dagger}=\sqrt{B(x)}-\sqrt{D(x)}\,e^{-\partial},
  \label{A,Ad}\\
  &\mathcal{A}_{x,y}=
  \sqrt{B(x)}\,\delta_{x,y}-\sqrt{D(x+1)}\,\delta_{x+1,y},\quad
  (\mathcal{A}^{\dagger})_{x,y}=
  \sqrt{B(x)}\,\delta_{x,y}-\sqrt{D(x)}\,\delta_{x-1,y}.
\end{align}
The zero mode $\mathcal{A}\phi_0(x)=0$ ($\phi_0(x)>0$) is easily obtained:
$\phi_0(x)^2=\prod_{y=0}^{x-1}\frac{B(y)}{D(y+1)}$,
with the normalization $\phi_0(0)=1$ (convention: $\prod_{k=n}^{n-1}*=1$).
We adopt the standard euclidean inner product $(\ \,,\ )$ of two real
functions on the grid as
$\bigl(f,g\bigr)\eqdef\sum_{x=0}^{x_{\text{max}}}f(x)g(x)$.
Then the orthogonality relation reads
\begin{equation}
  \bigl(\phi_n,\phi_m\bigr)
  =\frac{1}{d_n^2}\,\delta_{nm}\quad
  (n,m=0,1,\ldots,n_{\text{max}}).
  \label{ortho}
\end{equation}
Here $1/d_n^2$ is the normalization constant to be specified later.

Shape invariance, a sufficient condition for the exact solvability
\cite{os12,os4,os13,os14}, is realised by specific dependence of the
potential functions on a set of parameters
$\bm{\lambda}=(\lambda_1,\lambda_2,\ldots)$,
to be denoted by $B(x;\bm{\lambda})$, $D(x;\bm{\lambda})$,
$\mathcal{A}(\bm{\lambda})$, $\mathcal{H}(\bm{\lambda})$,
$\mathcal{E}_n(\bm{\lambda})$, $\phi_n(x;\bm{\lambda})$ etc.
The shape invariance condition is
\begin{equation}
  \mathcal{A}(\bm{\lambda})\mathcal{A}(\bm{\lambda})^{\dagger}
  =\kappa\mathcal{A}(\bm{\lambda}+\bm{\delta})^{\dagger}
  \mathcal{A}(\bm{\lambda}+\bm{\delta})
  +\mathcal{E}_1(\bm{\lambda}),
  \label{shapeinv1}
\end{equation}
where $\bm{\delta}$ is a certain shift of parameters and $\kappa$ is
a positive constant.
It should be stressed that the above definition is much stronger than
the original definition by Gendenshtein \cite{genden}.
The shape invariance condition \eqref{shapeinv1} combined with the Crum's
theorem \cite{crum,os15,gos,os22} implies that the entire energy spectrum
and the excited states eigenfunctions are expressed in terms of
$\mathcal{E}_1(\bm{\lambda})$ and $\phi_0(x;\bm{\lambda})$ as follows:
\begin{align}
  &\mathcal{E}_n(\bm{\lambda})=\sum_{s=0}^{n-1}
  \kappa^s\mathcal{E}_1(\bm{\lambda}+s\bm{\delta}),\\
  &\phi_n(x;\bm{\lambda})\propto
  \mathcal{A}(\bm{\lambda})^{\dagger}
  \mathcal{A}(\bm{\lambda}+\bm{\delta})^{\dagger}
  \mathcal{A}(\bm{\lambda}+2\bm{\delta})^{\dagger}
  \cdots
  \mathcal{A}(\bm{\lambda}+(n-1)\bm{\delta})^{\dagger}
  \phi_0(x;\bm{\lambda}+n\bm{\delta}).
\end{align}
We have also
\begin{align}
  \mathcal{A}(\bm{\lambda})\phi_n(x;\bm{\lambda})
  &=\frac{1}{\sqrt{B(0;\bm{\lambda})}}\,f_n(\bm{\lambda})
  \phi_{n-1}\bigl(x;\bm{\lambda}+\bm{\delta}\bigr),
  \label{Aphi=fphi}\\
  \mathcal{A}(\bm{\lambda})^{\dagger}
  \phi_{n-1}\bigl(x;\bm{\lambda}+\bm{\delta}\bigr)
  &=\sqrt{B(0;\bm{\lambda})}\,b_{n-1}(\bm{\lambda})\phi_n(x;\bm{\lambda}),
  \label{Adphi=bphi}
\end{align}
where $f_n(\bm{\lambda})$ and $b_{n-1}(\bm{\lambda})$ are the factors of
the energy eigenvalue,
$\mathcal{E}_n(\bm{\lambda})=f_n(\bm{\lambda})b_{n-1}(\bm{\lambda})$.

For the ($q$)-Racah and the other polynomials to be discussed in the
present paper, the eigenfunction has the following factorised form,
\begin{equation}
  \phi_n(x;\bm{\lambda})
  =\phi_0(x;\bm{\lambda})\check{P}_n(x;\bm{\lambda}),\quad
  \check{P}_n(x;\bm{\lambda})\eqdef
  P_n(\eta(x;\bm{\lambda});\bm{\lambda}),
  \label{phin=phi0Pn}
\end{equation}
where $P_n(\eta(x;\bm{\lambda});\bm{\lambda})$ is a polynomial of degree
$n$ in the sinusoidal coordinate $\eta(x;\bm{\lambda})$.
The sinusoidal coordinate considered here is a monotone increasing
function of $x$ satisfying the boundary condition
$\eta(0;\bm{\lambda})=0$ \cite{os12,os14}.
We choose the normalization
\begin{equation}
  P_n(0;\bm{\lambda})=1,
  \label{P(0)=1}
\end{equation}
and set $\check{P}_{-1}(x;\bm{\lambda})=0$. For later convenience,
let us remark on the relation
\begin{equation}
  P_n(\eta(1;\bm{\lambda});\bm{\lambda})= \check{P}_n(1;\bm{\lambda})
  =1-\frac{\mathcal{E}_n(\bm{\lambda})}{B(0;\bm{\lambda})}.
  \label{Pn1}
\end{equation}
The orthogonality relation \eqref{ortho} becomes
\begin{equation}
  \sum_{x=0}^{x_{\text{max}}}\phi_0(x;\bm{\lambda})^2
  \check{P}_n(x;\bm{\lambda})\check{P}_m(x;\bm{\lambda})
  =\frac{1}{d_n(\bm{\lambda})^2}\,\delta_{nm}\quad
  (n,m=0,1,\ldots,n_{\text{max}}).
\end{equation}
The forward shift operator $\mathcal{F}(\bm{\lambda})
=(\mathcal{F}_{x,y}(\bm{\lambda}))$,
the backward shift operator $\mathcal{B}(\bm{\lambda})
=(\mathcal{B}_{x,y}(\bm{\lambda}))$
and the similarity transformed Hamiltonian
$\widetilde{\mathcal{H}}(\bm{\lambda})
=(\widetilde{\mathcal{H}}_{x,y}(\bm{\lambda}))$
($x,y=0,1,\ldots,x_{\text{max}}$) are defined by
\begin{align}
  \mathcal{F}(\bm{\lambda})&\eqdef
  \sqrt{B(0;\bm{\lambda})}\,\phi_0(x;\bm{\lambda}+\bm{\delta})^{-1}\circ
  \mathcal{A}(\bm{\lambda})\circ\phi_0(x;\bm{\lambda})\n
  &=B(0;\bm{\lambda})\varphi(x;\bm{\lambda})^{-1}(1-e^{\partial}),
  \label{Fdef}\\
  \mathcal{B}(\bm{\lambda})&\eqdef
  \frac{1}{\sqrt{B(0;\bm{\lambda})}}\,\phi_0(x;\bm{\lambda})^{-1}\circ
  \mathcal{A}(\bm{\lambda})^{\dagger}
  \circ\phi_0(x;\bm{\lambda}+\bm{\delta})\n
  &=\frac{1}{B(0;\bm{\lambda})}\bigl(B(x;\bm{\lambda})
  -D(x;\bm{\lambda})e^{-\partial}\bigr)\varphi(x;\bm{\lambda}),
  \label{Bdef}\\
  \widetilde{\mathcal{H}}(\bm{\lambda})&\eqdef
  \phi_0(x;\bm{\lambda})^{-1}\circ\mathcal{H}(\bm{\lambda})
  \circ\phi_0(x;\bm{\lambda})
  =\mathcal{B}(\bm{\lambda})\mathcal{F}(\bm{\lambda})\n
  &=B(x;\bm{\lambda})(1-e^{\partial})+D(x;\bm{\lambda})(1-e^{-\partial}),
\end{align}
where the auxiliary functions $\varphi(x)$ is defined by \cite{os12}
\begin{equation}
  \varphi(x;\bm{\lambda})\eqdef
  \sqrt{\frac{B(0;\bm{\lambda})}{B(x;\bm{\lambda})}}
  \frac{\phi_0(x;\bm{\lambda}+\bm{\delta})}{\phi_0(x;\bm{\lambda})}
  =\frac{\eta(x+1;\bm{\lambda})-\eta(x;\bm{\lambda})}{\eta(1;\bm{\lambda})},
  \quad  \varphi(0;\bm{\lambda})=1.
  \label{phidef}
\end{equation}
Their action on the polynomials is ($n=0,1,\ldots,n_{\text{max}}$)
\begin{align}
  \mathcal{F}(\bm{\lambda})\check{P}_n(x;\bm{\lambda})
  &=f_n(\bm{\lambda})\check{P}_{n-1}(x;\bm{\lambda}+\bm{\delta}),
  \label{forwardrel}\\
  \mathcal{B}(\bm{\lambda})\check{P}_{n-1}(x;\bm{\lambda}+\bm{\delta})
  &=b_{n-1}(\bm{\lambda})\check{P}_n(x;\bm{\lambda}),
  \label{backwardrel}\\
  \widetilde{\mathcal{H}}(\bm{\lambda})\check{P}_n(x;\bm{\lambda})
  &=\mathcal{E}_n(\bm{\lambda})\check{P}_n(x;\bm{\lambda}).
  \label{tHPn=EPn}
\end{align}
The above difference equation \eqref{tHPn=EPn} for the polynomial $P_n$
reads explicitly as
\begin{align}
  B(x)\bigl(P_n(\eta(x))-P_n(\eta(x+1))\bigr)
  +D(x)\bigl(P_n(\eta(x))-P_n(\eta(x-1))\bigr)=\mathcal{E}_nP_n(\eta(x)),
  \label{difeqP}
\end{align}
in which the parameter dependence is suppressed for simplicity.

%%%%%%%%%%%%%%%%%%%%%%%%%%%%%%%%%%%%%%%%%%%%%%%%%%%%%%%%%%%%%%%
%                                                             %
%  3. Orignal System: ($q$-)Racah polynomials                 %
%                                                             %
%%%%%%%%%%%%%%%%%%%%%%%%%%%%%%%%%%%%%%%%%%%%%%%%%%%%%%%%%%%%%%%
\section{Original Systems: ($q$)-Racah polynomials}
\label{sec:original}
\setcounter{equation}{0}

Here we present various properties of the Racah (R) and the $q$-Racah
($q$R) polynomials as explored in \cite{os12}.
In general there are four cases of possible parameter choices indexed by
$(\epsilon,\epsilon')=(\pm 1,\pm 1)$. Here we restrict ourselves to the
$(\epsilon,\epsilon')=(1,1)$ case for simplicity of presentation.

The set of parameters $\bm{\lambda}$, which is different from the
standard one $(\alpha,\beta,\gamma,\delta)$ \cite{koeswart}, its shift
$\bm{\delta}$ and $\kappa$ are
\begin{align}
  \text{R}&:\ \bm{\lambda\,}=(a,b,c,d),\quad \bm{\delta}=(1,1,1,1),
  \quad\kappa=1,\\
  \text{$q$R}&:\ q^{\bm{\lambda}}=(a,b,c,d),\quad \bm{\delta}=(1,1,1,1),
  \quad\kappa=q^{-1}, \quad 0<q<1,
\end{align}
where $q^{\bm{\lambda}}$ stands for
$q^{(\lambda_1,\lambda_2,\ldots)}=(q^{\lambda_1},q^{\lambda_2},\ldots)$.
We introduce a new parameter  $\tilde{d}$ defined by
\begin{equation}
  \tilde{d}\eqdef
  \left\{
  \begin{array}{ll}
  a+b+c-d-1&:\text{R}\\
  abcd^{-1}q^{-1}&:\text{$q$R}
  \end{array}\right..
\end{equation}
The Hamiltonian is a finite dimensional matrix and the maximal values of
$x$ and $n$ are $x_{\text{max}}=n_{\text{max}}=N$ and we could choose
\begin{alignat}{2}
  &\text{R}\ &:\ \ &a=-N\quad\text{or}\quad b=-N\quad\text{or}\quad c=-N,\n
  &\text{$q$R}\ &:\ \ &a=q^{-N}\quad\text{or}\quad b=q^{-N}\quad\text{or}
  \quad c=q^{-N},
  \label{a,b,c}
\end{alignat}
to ensures the boundary condition for $B$, $B(x_{\text{max}})=0$.
The potential functions $B(x;\bm{\lambda})$ and $D(x;\bm{\lambda})$ are
\begin{align}
  &B(x;\bm{\lambda})=
  \left\{
  \begin{array}{ll}
  {\displaystyle
  -\frac{(x+a)(x+b)(x+c)(x+d)}{(2x+d)(2x+1+d)}}&:\text{R}\\[8pt]
  {\displaystyle-\frac{(1-aq^x)(1-bq^x)(1-cq^x)(1-dq^x)}
  {(1-dq^{2x})(1-dq^{2x+1})}}&:\text{$q$R}
  \end{array}\right.\!,\\
  &D(x;\bm{\lambda})=
  \left\{
  \begin{array}{ll}
  {\displaystyle
  -\frac{(x+d-a)(x+d-b)(x+d-c)x}{(2x-1+d)(2x+d)}}&:\text{R}\\[8pt]
  {\displaystyle-\tilde{d}\,
  \frac{(1-a^{-1}dq^x)(1-b^{-1}dq^x)(1-c^{-1}dq^x)(1-q^x)}
  {(1-dq^{2x-1})(1-dq^{2x})}}&:\text{$q$R}
  \end{array}\right.\!.
\end{align}
The parameter ranges are restricted by the positivity of
$B(x;\bm{\lambda})$ and $D(x;\bm{\lambda})$.
When we need to specify them, we adopt the following choice of the
parameter ranges:
\begin{alignat}{2}
  &\text{R}\ &:\ \ &a=-N,\ \ a+b>d>0,\ \ 0<c<1+d,\n
  &\text{$q$R}\ &:\ \ &  a=q^{-N},\ \ 0<ab<d<1,\ \ qd<c<1.
  \label{para_range}
\end{alignat}
The energy eigenvalue and the sinusoidal coordinate are
\begin{equation}
  \mathcal{E}_n(\bm{\lambda})=
  \left\{
  \begin{array}{ll}
  n(n+\tilde{d})&:\text{R}\\
  (q^{-n}-1)(1-\tilde{d}q^n)&:\text{$q$R}
  \end{array}\right.\!,\quad
  \eta(x;\bm{\lambda})=
  \left\{
  \begin{array}{ll}
  x(x+d)&:\text{R}\\
  (q^{-x}-1)(1-dq^x)&:\text{$q$R}
  \end{array}\right.\!.
\end{equation}
The eigenfunctions have the factorised form \eqref{phin=phi0Pn} and the
orthogonal polynomials are the Racah and the $q$-Racah polynomials:
\begin{align}
  \check{P}_n(x;\bm{\lambda})=P_n(\eta(x;\bm{\lambda});\bm{\lambda})&=
  \left\{
  \begin{array}{ll}
  {\displaystyle
  {}_4F_3\Bigl(
  \genfrac{}{}{0pt}{}{-n,\,n+\tilde{d},\,-x,\,x+d}
  {a,\,b,\,c}\Bigm|1\Bigr)}&:\text{R}\\
  {\displaystyle
  {}_4\phi_3\Bigl(
  \genfrac{}{}{0pt}{}{q^{-n},\,\tilde{d}q^n,\,q^{-x},\,dq^x}
  {a,\,b,\,c}\Bigm|q\,;q\Bigr)}&:\text{$q$R}
  \end{array}\right.\\
  &=\left\{
  \begin{array}{ll}
  {\displaystyle
  R_n(\eta(x;\bm{\lambda});a-1,\tilde{d}-a,c-1,d-c)}
  &:\text{R}\\
  {\displaystyle
  R_n(1+d+\eta(x;\bm{\lambda});
  aq^{-1},\tilde{d}a^{-1},cq^{-1},dc^{-1}|q)}&:\text{$q$R}
  \end{array}\right.\!.
  \label{Pn=R,qR}
\end{align}
Here $R_n(\cdots)$ are the standard notation in \cite{koeswart}.
The auxiliary function $\varphi(x;\bm{\lambda})$ \eqref{phidef} reads
\begin{equation}
  \varphi(x;\bm{\lambda})=
  \left\{
  \begin{array}{ll}
  {\displaystyle\frac{2x+d+1}{d+1}}&:\text{R}\\[6pt]
  {\displaystyle\frac{q^{-x}-dq^{x+1}}{1-dq}}&:\text{$q$R}
  \end{array}\right.\!.
\end{equation}
The constants $f_n(\bm{\lambda})$ and $b_n(\bm{\lambda})$ appearing in
\eqref{Aphi=fphi}--\eqref{Adphi=bphi} are
\begin{equation}
  f_n(\bm{\lambda})=\mathcal{E}_n(\bm{\lambda}),\quad
  b_n(\bm{\lambda})=1.
\end{equation}
The orthogonality measure $\phi_0(x;\bm{\lambda})^2$ and the
normalisation constants $d_n(\bm{\lambda})^2$ are
\begin{align}
  &\phi_0(x;\bm{\lambda})^2=\left\{
  \begin{array}{ll}
  {\displaystyle
  \frac{(a,b,c,d)_x}{(1+d-a,1+d-b,1+d-c,1)_x}\,
  \frac{2x+d}{d}
  }&:\text{R}\\[8pt]
  {\displaystyle
  \frac{(a,b,c,d\,;q)_x}
  {(a^{-1}dq,b^{-1}dq,c^{-1}dq,q\,;q)_x\,\tilde{d}^x}\,
  \frac{1-dq^{2x}}{1-d}
  }&:\text{$q$R}
  \end{array}\right.\!,\\
  &d_n(\bm{\lambda})^2
  =\left\{
  \begin{array}{ll}
  {\displaystyle
  \frac{(a,b,c,\tilde{d})_n}
  {(1+\tilde{d}-a,1+\tilde{d}-b,1+\tilde{d}-c,1)_n}\,
  \frac{2n+\tilde{d}}{\tilde{d}}
  }&\\[8pt]
  {\displaystyle
  \quad\times
  \frac{(-1)^N(1+d-a,1+d-b,1+d-c)_N}{(\tilde{d}+1)_N(d+1)_{2N}}
  }&:\text{R}\\[8pt]
  {\displaystyle
  \frac{(a,b,c,\tilde{d}\,;q)_n}
  {(a^{-1}\tilde{d}q,b^{-1}\tilde{d}q,c^{-1}\tilde{d}q,q\,;q)_n\,d^n}\,
  \frac{1-\tilde{d}q^{2n}}{1-\tilde{d}}
  }&\\[8pt]
  {\displaystyle
  \quad\times
  \frac{(-1)^N(a^{-1}dq,b^{-1}dq,c^{-1}dq\,;q)_N\,\tilde{d}^Nq^{\frac12N(N+1)}}
  {(\tilde{d}q\,;q)_N(dq\,;q)_{2N}}
  }&:\text{$q$R}
  \end{array}\right.\!.
  \label{dn2}
\end{align}

%%%%%%%%%%%%%%%%%%%%%%%%%%%%%%%%%%%%%%%%%%%%%%%%%%%%%%%%%%%%%%%
%                                                             %
%  4. Deformed System: $X_{\ell}$ ($q$-)Racah polynomials     %
%                                                             %
%%%%%%%%%%%%%%%%%%%%%%%%%%%%%%%%%%%%%%%%%%%%%%%%%%%%%%%%%%%%%%%
\section{Deformed Systems: $X_{\ell}$ ($q$)-Racah polynomials}
\label{sec:deformed}
\setcounter{equation}{0}

For each positive integer $\ell=1,2,\ldots, N-1$, we can construct a shape
invariant system by deforming the original system ($\ell=0$) in terms of
a degree $\ell$ eigenpolynomial $\xi_{\ell}$ of twisted parameters.

We set
\begin{equation}
  x_{\text{max}}^{\ell}\eqdef N-\ell,\quad
  n_{\text{max}}^{\ell}\eqdef N-\ell,
\end{equation}
and take
\begin{alignat}{2}
  &\text{R}\ &:\ \ &a=-N\quad\text{or}\quad b=-N,\n
  &\text{$q$R}\ &:\ \ &a=q^{-N}\quad\text{or}\quad b=q^{-N}.
  \label{a,b}
\end{alignat}

The deforming polynomial $\xi_{\ell}$, which is a polynomial of degree
$\ell$ in $\eta(x;\bm{\lambda}+(\ell-1)\bm{\delta})$, is defined
from the eigenpolynomial $\check{P}_{\ell}(x;\bm{\lambda})$:
\begin{align}
  \check{\xi}_{\ell}(x;\bm{\lambda})
  &\eqdef\xi_{\ell}(\eta(x;\bm{\lambda}+(\ell-1)\bm{\delta});\bm{\lambda})\n
  &\eqdef\check{P}_{\ell}\bigl(x;\mathfrak{t}
  \bigl(\bm{\lambda}+(\ell-1)\bm{\delta}\bigr)\bigr)
  \ :\,\text{R,\,$q$R}\n
  &=\left\{
  \begin{array}{ll}
  {\displaystyle
  {}_4F_3\Bigl(
  \genfrac{}{}{0pt}{}{-\ell,\,\ell-a-b+c+d-1,\,-x,\,x+d+\ell-1}
  {d-a,\,d-b,\,c+\ell-1}\Bigm|1\Bigr)}&:\text{R}\\
  {\displaystyle
  {}_4\phi_3\Bigl(
  \genfrac{}{}{0pt}{}
  {q^{-\ell},\,a^{-1}b^{-1}cdq^{\ell-1},\,q^{-x},\,dq^{x+\ell-1}}
  {a^{-1}d,\,b^{-1}d,\,cq^{\ell-1}}\Bigm|q\,;q\Bigr)}&:\text{$q$R}
  \end{array}\right.\!,
\end{align}
which satisfies the normalization
\begin{equation}
  \xi_{\ell}(0;\bm{\lambda})=1.
\end{equation}
Here the twist operator $\mathfrak{t}$ acting on the set of
parameters $\bm{\lambda}=(\lambda_1,\lambda_2,\lambda_3,\lambda_4)$ is
\begin{equation}
  \mathfrak{t}(\bm{\lambda})\eqdef
  (\lambda_4-\lambda_1,\lambda_4-\lambda_2,\lambda_3,\lambda_4)
  \ :\,\text{R,\,$q$R}.
\end{equation}
This is the most important ingredient of the deformation.
For the appropriate parameter ranges, for example as given in
\eqref{para_range}, the deforming polynomial
$\check{\xi}_{\ell}(x;\bm{\lambda})$ is positive at integer points
$x=0,1,\ldots,x_{\text{max}}^{\ell}+1$, because the polynomial
$\xi_{\ell}(y;\bm{\lambda})$ has no zeros in the interval
$0\leq y\leq\eta(x_{\text{max}}^{\ell}+1;\bm{\lambda}+(\ell-1)\bm{\delta})$.
It satisfies the following two formulas,
which will play important roles in the derivation of various results:
\begin{align}
  &\frac{1}{\varphi(x;\bm{\lambda}+\ell\bm{\delta}+\tilde{\bm{\delta}})}
  \Bigl(
  v_1^B(x;\bm{\lambda}+\ell\bm{\delta})
  -v_1^D(x;\bm{\lambda}+\ell\bm{\delta})e^{\partial}\Bigr)
  \check{\xi}_{\ell}(x;\bm{\lambda})
  =\hat{f}_{\ell,0}(\bm{\lambda})
  \check{\xi}_{\ell}(x;\bm{\lambda}+\bm{\delta}),
  \label{xil(l+d)}\\
  &\frac{1}{\varphi(x;\bm{\lambda}+(\ell-1)\bm{\delta}+\tilde{\bm{\delta}})}
  \Bigl(
  v_2^B(x;\bm{\lambda}+(\ell-1)\bm{\delta})
  -v_2^D(x;\bm{\lambda}+(\ell-1)\bm{\delta})e^{-\partial}\Bigr)
  \check{\xi}_{\ell}(x;\bm{\lambda}+\bm{\delta})\n
  &\qquad\quad
  =\hat{b}_{\ell,0}(\bm{\lambda})\check{\xi}_{\ell}(x;\bm{\lambda}).
  \label{xil(l)}
\end{align}
Here $v_1^B(x;\bm{\lambda})$, $v_2^B(x;\bm{\lambda})$,
$v_1^D(x;\bm{\lambda})$, $v_2^D(x;\bm{\lambda})$ are the factors of
the potential functions $B(x;\bm{\lambda})$ and $D(x;\bm{\lambda})$:
\begin{align}
  v_1^B(x;\bm{\lambda})&\eqdef
  \left\{
  \begin{array}{ll}
  d^{-1}(x+a)(x+b)&:\text{R}\\
  {\displaystyle
  \frac{q^{-x}}{1-d}(1-aq^x)(1-bq^x)
  }&:\text{$q$R}
  \end{array}\right.\!,
  \label{v1B}\\
  v_2^B(x;\bm{\lambda})&\eqdef
  \left\{
  \begin{array}{ll}
  d^{-1}(x+c)(x+d)&:\text{R}\\
  {\displaystyle
  \frac{q^{-x}}{1-d}(1-cq^x)(1-dq^x)
  }&:\text{$q$R}
  \end{array}\right.\!,
  \label{v2B}\\
  v_1^D(x;\bm{\lambda})&\eqdef
  \left\{
  \begin{array}{ll}
  d^{-1}(x+d-a)(x+d-b)&:\text{R}\\
  {\displaystyle
  \frac{q^{-x}}{1-d}abd^{-1}(1-a^{-1}dq^x)(1-b^{-1}dq^x)
  }&:\text{$q$R}
  \end{array}\right.\!,
  \label{v1D}\\
  v_2^D(x;\bm{\lambda})&\eqdef
  \left\{
  \begin{array}{ll}
  d^{-1}(x+d-c)x&:\text{R}\\
  {\displaystyle
  \frac{q^{-x}}{1-d}c(1-c^{-1}dq^x)(1-q^x)
  }&:\text{$q$R}
  \end{array}\right.\!,
  \label{v2D}\\
  B(x;\bm{\lambda})&=-\sqrt{\kappa}\,
  \frac{v_1^B(x;\bm{\lambda})v_2^B(x;\bm{\lambda})}
  {\varphi(x;\bm{\lambda}+\tilde{\bm{\delta}})
  \varphi(x+\frac12;\bm{\lambda}+\tilde{\bm{\delta}})},
  \label{factorB}\\
  D(x;\bm{\lambda})&=-\sqrt{\kappa}\,
  \frac{v_1^D(x;\bm{\lambda})v_2^D(x;\bm{\lambda})}
  {\varphi(x;\bm{\lambda}+\tilde{\bm{\delta}})
   \varphi(x-\frac12;\bm{\lambda}+\tilde{\bm{\delta}})},
  \label{factorD}
\end{align}
where $\tilde{\bm{\delta}}$ is
\begin{equation}
  \tilde{\bm{\delta}}\eqdef(0,0,-1,-1)\ :\,\text{R, $q$R}.
\end{equation}
The constants $\hat{f}_{\ell,n}(\bm{\lambda})$ and
$\hat{b}_{\ell,n}(\bm{\lambda})$ are given by
\begin{equation}
  \hat{f}_{\ell,n}(\bm{\lambda})\eqdef
  \left\{
  \begin{array}{ll}
  {\displaystyle
  (a+b-d+n)\frac{c+2\ell+n-1}{c+\ell-1}
  }&:\text{R}\\
  {\displaystyle
  q^{-n}(1-abd^{-1}q^n)\frac{1-cq^{2\ell+n-1}}{1-cq^{\ell-1}}
  }&:\text{$q$R}
  \end{array}\right.\!,\quad
  \hat{b}_{\ell,n}(\bm{\lambda})\eqdef
  \left\{
  \begin{array}{ll}
  c+\ell-1&:\text{R}\\
  1-cq^{\ell-1}&:\text{$q$R}
  \end{array}\right.\!.
  \label{fhlnbhln}
\end{equation}
The eqs.\,\eqref{xil(l+d)}--\eqref{xil(l)} are identities relating
$\check{\xi}_{\ell}(x;\bm{\lambda})$ and
$\check{\xi}_{\ell}(x;\bm{\lambda}+\bm{\delta})$.
They are reduced to the identities satisfied by the (basic)
hypergeometric functions, (2.74)--(2.75) in \cite{os20}.
Note that these two equations \eqref{xil(l+d)}--\eqref{xil(l)}
imply the difference equation for the deforming polynomial,
\begin{equation}
  \Bigl(B\bigl(x;\mathfrak{t}(\bm{\lambda}+(\ell-1)\bm{\delta})\bigr)
  (1-e^{\partial})
  +D\bigl(x;\mathfrak{t}(\bm{\lambda}+(\ell-1)\bm{\delta})\bigr)
  (1-e^{-\partial})\Bigr)\check{\xi}_{\ell}(x;\bm{\lambda})
  =\mathcal{E}_{\ell}(\mathfrak{t}(\bm{\lambda}))
  \check{\xi}_{\ell}(x;\bm{\lambda}),
  \label{xildiffeq}
\end{equation}
which corresponds to \eqref{tHPn=EPn}.

Let us introduce new potential functions
$B_{\ell}(x;\bm{\lambda})$ and
$D_{\ell}(x;\bm{\lambda})$ by multiplicatively deforming the original ones
in terms of the  polynomial $\check{\xi}_{\ell}(x;\bm{\lambda})$:
\begin{align}
  &B_{\ell}(x;\bm{\lambda})\eqdef B(x;\bm{\lambda}+\ell\bm{\delta})
  \frac{\check{\xi}_{\ell}(x;\bm{\lambda})}
  {\check{\xi}_{\ell}(x+1;\bm{\lambda})}
  \frac{\check{\xi}_{\ell}(x+1;\bm{\lambda}+\bm{\delta})}
  {\check{\xi}_{\ell}(x;\bm{\lambda}+\bm{\delta})},
  \label{Bl}\\
  &D_{\ell}(x;\bm{\lambda})\eqdef D(x;\bm{\lambda}+\ell\bm{\delta})
  \frac{\check{\xi}_{\ell}(x+1;\bm{\lambda})}
  {\check{\xi}_{\ell}(x;\bm{\lambda})}
  \frac{\check{\xi}_{\ell}(x-1;\bm{\lambda}+\bm{\delta})}
  {\check{\xi}_{\ell}(x;\bm{\lambda}+\bm{\delta})}.
  \label{Dl}
\end{align}
See the corresponding expressions for the exceptional Wilson and Askey-Wilson
polynomials (30)--(31) of \cite{os17} and (2.42)--(2.43) of \cite{os20}.
They define a deformed Hamiltonian
$\mathcal{H}_{\ell}(\bm{\lambda})=(\mathcal{H}_{\ell;x,y}(\bm{\lambda}))$
and other operators
$\mathcal{A}_{\ell}(\bm{\lambda})=(\mathcal{A}_{\ell;x,y}(\bm{\lambda}))$
and $\mathcal{A}_{\ell}(\bm{\lambda})^{\dagger}
=((\mathcal{A}_{\ell}(\bm{\lambda})^{\dagger})_{x,y})
=(\mathcal{A}_{\ell;y,x}(\bm{\lambda}))$
$(x,y=0,1,$ $\ldots,x_{\text{max}}^{\ell}$) by
\begin{align}
  &\mathcal{H}_{\ell}(\bm{\lambda})
  \eqdef\mathcal{A}_{\ell}(\bm{\lambda})^{\dagger}
  \mathcal{A}_{\ell}(\bm{\lambda}),\\
  &\mathcal{A}_{\ell}(\bm{\lambda})
  \eqdef\sqrt{B_{\ell}(x;\bm{\lambda})}
  -e^{\partial}\sqrt{D_{\ell}(x;\bm{\lambda})},\quad
  \mathcal{A}_{\ell}(\bm{\lambda})^{\dagger}
  =\sqrt{B_{\ell}(x;\bm{\lambda})}
  -\sqrt{D_{\ell}(x;\bm{\lambda})}\,e^{-\partial}.
  \label{Al,Ald}
\end{align}
 We have $D_{\ell}(0;\bm{\lambda})=0$ and
$B_{\ell}(x_{\text{max}}^{\ell};\bm{\lambda})=0$.
The parameter ranges are restricted by the positivity of
$B_{\ell}(x;\bm{\lambda})$ and $D_{\ell}(x;\bm{\lambda})$.
When we need to specify them, we consider the parameter ranges
\eqref{para_range}.

The deformed system is shape invariant, too:
\begin{equation}
  \mathcal{A}_{\ell}(\bm{\lambda})\mathcal{A}_{\ell}(\bm{\lambda})^{\dagger}
  =\kappa\mathcal{A}_{\ell}(\bm{\lambda}+\bm{\delta})^{\dagger}
  \mathcal{A}_{\ell}(\bm{\lambda}+\bm{\delta})
  +\mathcal{E}_{\ell,1}(\bm{\lambda}),
  \label{shapeinvl1}
\end{equation}
or equivalently,
\begin{align}
  \sqrt{B_{\ell}(x+1;\bm{\lambda})D_{\ell}(x+1;\bm{\lambda})}
  &=\kappa\sqrt{B_{\ell}(x;\bm{\lambda}+\bm{\delta})
  D_{\ell}(x+1;\bm{\lambda}+\bm{\delta})},
  \label{shapeinvl1cond1}\\
  B_{\ell}(x;\bm{\lambda})+D_{\ell}(x+1;\bm{\lambda})
  &=\kappa\bigl(B_{\ell}(x;\bm{\lambda}+\bm{\delta})
  +D_{\ell}(x;\bm{\lambda}+\bm{\delta})
  \bigr)+\mathcal{E}_{\ell,1}(\bm{\lambda}).
  \label{shapeinvl1cond2}
\end{align}
The proof is straightforward by direct calculation. In order to verify
\eqref{shapeinvl1cond2}, use is made of the two properties of the
deforming polynomial $\check{\xi}_{\ell}(x;\bm{\lambda})$
\eqref{xil(l+d)}--\eqref{xil(l)}.

The Schr\"{o}dinger equation of the modified system is
($n=0,1,\ldots,n_{\text{max}}^{\ell}$)
\begin{equation}
  \mathcal{H}_{\ell}(\bm{\lambda})\phi_{\ell,n}(x;\bm{\lambda})
  =\mathcal{E}_{\ell,n}(\bm{\lambda})\phi_{\ell,n}(x;\bm{\lambda}),\quad
  \mathcal{E}_{\ell,n}(\bm{\lambda})
  \eqdef\mathcal{E}_n(\bm{\lambda}+\ell\bm{\delta}).
\end{equation}
The ground state $\phi_{\ell,0}(x;\bm{\lambda})$, which is annihilated by
$\mathcal{A}_{\ell}(\bm{\lambda})$, is
\begin{align}
  &\phi_{\ell,0}(x;\bm{\lambda})=\sqrt{\prod_{y=0}^{x-1}
  \frac{B_{\ell}(y;\bm{\lambda})}{D_{\ell}(y+1;\bm{\lambda})}}
  =\psi_{\ell}(x;\bm{\lambda})\check{\xi}_{\ell}(x;\bm{\lambda}+\bm{\delta}),
  \label{phil0}\\
  &\psi_{\ell}(x;\bm{\lambda})\eqdef\phi_0(x;\bm{\lambda}+\ell\bm{\delta})
  \sqrt{\frac{\check{\xi}_{\ell}(1;\bm{\lambda})}
  {\check{\xi}_{\ell}(x;\bm{\lambda})\check{\xi}_{\ell}(x+1;\bm{\lambda})}},
\end{align}
with the normalisation $\phi_{\ell,0}(0;\bm{\lambda})=1$ and
$\psi_{\ell}(0;\bm{\lambda})=1$.
The excited states wavefunctions have the factorised form as
\eqref{phin=phi0Pn}:
\begin{equation}
  \phi_{\ell,n}(x;\bm{\lambda})=\psi_{\ell}(x;\bm{\lambda})
  \check{P}_{\ell,n}(x;\bm{\lambda}).
\end{equation}
The {\em exceptional\/} ($X_\ell$) ($q$)-{\em Racah polynomial\/}
$\check{P}_{\ell,n}(x;\bm{\lambda})$ is bilinear in the deforming
polynomial $\check{\xi}_\ell$ and the original polynomial $\check{P}_n$:
\begin{align}
  &\check{P}_{\ell,n}(x;\bm{\lambda})
  \eqdef P_{\ell,n}(\eta(x;\bm{\lambda}+\ell\bm{\delta});\bm{\lambda})\n
  &\eqdef\frac{1}{\hat{f}_{\ell,n}(\bm{\lambda})}
  \frac{1}{\varphi(x;\bm{\lambda}+\ell\bm{\delta}+\tilde{\bm{\delta}})}
  \Bigl(v_1^B(x;\bm{\lambda}+\ell\bm{\delta})
  \check{\xi}_{\ell}(x;\bm{\lambda})
  \check{P}_{n}(x+1;\bm{\lambda}+\ell\bm{\delta}+\tilde{\bm{\delta}})\n
  &\qquad\qquad\qquad\qquad\qquad\qquad
  -v_1^D(x;\bm{\lambda}+\ell\bm{\delta})\check{\xi}_{\ell}(x+1;\bm{\lambda})
  \check{P}_{n}(x;\bm{\lambda}+\ell\bm{\delta}+\tilde{\bm{\delta}})\Bigr).
  \label{Pln}
\end{align}
This is one of the main results of the present paper to be compared with
the similar expressions for the exceptional Laguerre \& Jacobi polynomials
(2.1)--(2.4) in \cite{hos}, (2.31),(2.33) \& (3.37),(3.40) of \cite{stz},
for the exceptional Wilson \& Askey-Wilson polynomials (2.52) in \cite{os20}.
The overall multiplicative factor is so chosen and as to realise the
normalisation condition
\begin{equation}
  P_{\ell,n}(0;\bm{\lambda})=1,
\end{equation}
which can be shown by using \eqref{Pn1}.
This is a polynomial of degree $\ell+n$ in
$\eta(x;\bm{\lambda}+\ell\bm{\delta})$. 
Note that $\check{P}_{\ell,0}(x;\bm{\lambda})
=\check{\xi}_{\ell}(x;\bm{\lambda}+\bm{\delta})$ due to \eqref{xil(l+d)},
which is obviously a polynomial of degree $\ell$ in
$\eta(x;\bm{\lambda}+\ell\bm{\delta})$. 
The exceptional orthogonal polynomial $P_{\ell,n}(y;\bm{\lambda})$ has $n$
real zeros in the interval
$0\leq y\leq\eta(x_{\text{max}}^{\ell};\bm{\lambda}+\ell\bm{\delta})$ for
the appropriate parameter ranges, for example the range \eqref{para_range}.
It has $\ell$ extra zeros which are usually complex and lie outside the
above interval.

The action of the operators $\mathcal{A}_{\ell}(\bm{\lambda})$ and
$\mathcal{A}_{\ell}(\bm{\lambda})^{\dagger}$ on the eigenfunctions is
\begin{align}
  \mathcal{A}_{\ell}(\bm{\lambda})\phi_{\ell,n}(x;\bm{\lambda})
  &=\frac{1}{\sqrt{B_{\ell}(0;\bm{\lambda})}}\,f_{\ell,n}(\bm{\lambda})
  \phi_{\ell,n-1}\bigl(x;\bm{\lambda}+\bm{\delta}\bigr),
  \label{Alphiln=flnphiln}\\
  \mathcal{A}_{\ell}(\bm{\lambda})^{\dagger}
  \phi_{\ell,n-1}\bigl(x;\bm{\lambda}+\bm{\delta}\bigr)
  &=\sqrt{B_{\ell}(0;\bm{\lambda})}\,
  b_{\ell,n-1}(\bm{\lambda})\phi_{\ell,n}(x;\bm{\lambda}),
  \label{Aldphiln=blnphiln}\\
  \quad f_{\ell,n}(\bm{\lambda})=f_n(\bm{\lambda}+\ell\bm{\delta}),
  &\quad b_{\ell,n-1}(\bm{\lambda})=b_{n-1}(\bm{\lambda}+\ell\bm{\delta}).
  \label{fln,bln}
\end{align}
Like the corresponding formulas of the original systems
\eqref{Aphi=fphi}--\eqref{Adphi=bphi}, these are simple consequences of
the shape invariance and the normalisation of the eigenfunctions.
In the next section, we will derive these formulas through the
intertwining relations and without recourse to the  shape invariance of the
deformed system \eqref{shapeinvl1}.
The forward shift operator $\mathcal{F}_{\ell}(\bm{\lambda})
=(\mathcal{F}_{\ell;x,y}(\bm{\lambda}))$,
the backward shift operator $\mathcal{B}_{\ell}(\bm{\lambda})
=(\mathcal{B}_{\ell;x,y}(\bm{\lambda}))$
and the similarity transformed Hamiltonian
$\widetilde{\mathcal{H}}_{\ell}(\bm{\lambda})
=(\widetilde{\mathcal{H}}_{\ell;x,y}(\bm{\lambda}))$
($x,y=0,1,\ldots,x_{\text{max}}^{\ell}$) are defined by
\begin{align}
  \mathcal{F}_{\ell}(\bm{\lambda})&\eqdef\sqrt{B_{\ell}(0;\bm{\lambda})}\,
  \psi_{\ell}(x;\bm{\lambda}+\bm{\delta})^{-1}\circ
  \mathcal{A}_{\ell}(\bm{\lambda})\circ\psi_{\ell}(x;\bm{\lambda})\n
  &=\frac{B(0,\bm{\lambda}+\ell\bm{\delta})}
  {\varphi(x;\bm{\lambda}+\ell\bm{\delta})
  \check{\xi}_{\ell}(x+1;\bm{\lambda})}
  \Bigl(\check{\xi}_{\ell}(x+1;\bm{\lambda}+\bm{\delta})
  -\check{\xi}_{\ell}(x;\bm{\lambda}+\bm{\delta})e^{\partial}\Bigr),
  \label{Fldef}\\
  \mathcal{B}_{\ell}(\bm{\lambda})&\eqdef
  \frac{1}{\sqrt{B_{\ell}(0;\bm{\lambda})}}\,
  \psi_{\ell}(x;\bm{\lambda})^{-1}\circ
  \mathcal{A}_{\ell}(\bm{\lambda})^{\dagger}
  \circ\psi_{\ell}\bigl(x;\bm{\lambda}+\bm{\delta}\bigr)\n
  &=\frac{1}{B(0;\bm{\lambda}+\ell\bm{\delta})}
  \frac{1}{\check{\xi}_{\ell}(x;\bm{\lambda}+\bm{\delta})}\n
  &\quad\times
  \Bigl(B(x;\bm{\lambda}+\ell\bm{\delta})\check{\xi}_{\ell}(x;\bm{\lambda})
  -D(x;\bm{\lambda}+\ell\bm{\delta})\check{\xi}_{\ell}(x+1;\bm{\lambda})
  e^{-\partial}\Bigl)
  \varphi(x;\bm{\lambda}+\ell\bm{\delta}),
  \label{Bldef}\\
  \widetilde{\mathcal{H}}_{\ell}(\bm{\lambda})&\eqdef
  \psi_{\ell}(x;\bm{\lambda})^{-1}\circ
  \mathcal{H}_{\ell}(\bm{\lambda})\circ\psi_{\ell}(x;\bm{\lambda})
  =\mathcal{B}_{\ell}(\bm{\lambda})\mathcal{F}_{\ell}(\bm{\lambda})\n
  &=B(x;\bm{\lambda}+\ell\bm{\delta})
  \frac{\check{\xi}_{\ell}(x;\bm{\lambda})}
  {\check{\xi}_{\ell}(x+1;\bm{\lambda})}
  \Bigl(\frac{\check{\xi}_{\ell}(x+1;\bm{\lambda}+\bm{\delta})}
  {\check{\xi}_{\ell}(x;\bm{\lambda}+\bm{\delta})}-e^{\partial}\Bigr)\n
  &\quad+D(x;\bm{\lambda}+\ell\bm{\delta})
  \frac{\check{\xi}_{\ell}(x+1;\bm{\lambda})}
  {\check{\xi}_{\ell}(x;\bm{\lambda})}
  \Bigl(\frac{\check{\xi}_{\ell}(x-1;\bm{\lambda}+\bm{\delta})}
  {\check{\xi}_{\ell}(x;\bm{\lambda}+\bm{\delta})}-e^{-\partial}\Bigr).
\end{align}
Compare with the similar expressions for the $X_\ell$ Laguerre \& Jacobi
polynomials (3.2)--(3.5) in \cite{hos},
and for the $X_\ell$ Wilson \& Askey-Wilson polynomials (2.58)--(2.63)
in \cite{os20}.
Their action on the polynomials is ($n=0,1,\ldots,n_{\text{max}}^{\ell}$)
\begin{align}
  \mathcal{F}_{\ell}(\bm{\lambda})\check{P}_{\ell,n}(x;\bm{\lambda})
  &=f_{\ell,n}(\bm{\lambda})
  \check{P}_{\ell,n-1}(x;\bm{\lambda}+\bm{\delta}),
  \label{FlPln=flnPln}\\
  \mathcal{B}_{\ell}(\bm{\lambda})
  \check{P}_{\ell,n-1}(x;\bm{\lambda}+\bm{\delta})
  &=b_{\ell,n-1}(\bm{\lambda})\check{P}_{\ell,n}(x;\bm{\lambda}),
  \label{BlPln=blnPln}\\
  \widetilde{\mathcal{H}}_{\ell}(\bm{\lambda})
  \check{P}_{\ell,n}(x;\bm{\lambda})
  &=\mathcal{E}_{\ell,n}(\bm{\lambda})\check{P}_{\ell,n}(x;\bm{\lambda}),
  \quad
  \mathcal{E}_{\ell,n}(\bm{\lambda})
  =\mathcal{E}_n(\bm{\lambda}+\ell\bm{\delta}).
  \label{difeqPl}
\end{align}

The orthogonality relation is
\begin{equation}
  \sum_{x=0}^{x_{\text{max}}^{\ell}}
  \frac{\psi_{\ell}(x;\bm{\lambda})^2}{\check{\xi}_{\ell}(1;\bm{\lambda})}\,
  \check{P}_{\ell,n}(x;\bm{\lambda})\check{P}_{\ell,m}(x;\bm{\lambda})
  =\frac{\delta_{nm}}{d_{\ell,n}(\bm{\lambda})^2}\quad
  (n,m=0,1,\ldots,n_{\text{max}}^{\ell}).
\end{equation}
The normalisation constants $d_{\ell,n}(\bm{\lambda})^2$ are
\begin{equation}
  d_{\ell,n}(\bm{\lambda})^2
  =d_n(\bm{\lambda}+\ell\bm{\delta}+\tilde{\bm{\delta}})^2\,
  \frac{\hat{f}_{\ell,n}(\bm{\lambda})}{\hat{b}_{\ell,n}(\bm{\lambda})}
  \frac{1}{s_{\ell}(\bm{\lambda})}
  =d_n(\bm{\lambda}+\ell\bm{\delta})^2\,
  \frac{\hat{f}_{\ell,n}(\bm{\lambda})}{\hat{b}_{\ell,n}(\bm{\lambda})}
  \frac{\hat{b}_{0,n}(\bm{\lambda}+\ell\bm{\delta})}
  {\hat{f}_{0,n}(\bm{\lambda}+\ell\bm{\delta})}
  \frac{s_0(\bm{\lambda}+\ell\bm{\delta})}{s_{\ell}(\bm{\lambda})},
  \label{dln2}
\end{equation}
where $s_{\ell}(\bm{\lambda})$ is defined by
\begin{equation}
  s_{\ell}(\bm{\lambda})\eqdef
  \left\{
  \begin{array}{ll}
  {\displaystyle
  -\frac{(d-a)(d-b)}{(c+\ell-1)(d+\ell)}
  }&:\text{R}\\
  {\displaystyle
  -abd^{-1}q^{\ell}
  \frac{(1-a^{-1}d)(1-b^{-1}d)}{(1-cq^{\ell-1})(1-dq^{\ell})}
  }&:\text{$q$R}
  \end{array}\right.\!.
  \label{sl}
\end{equation}
This will be proved in the next section.
In the second equality of \eqref{dln2} use is made of the explicit forms
of $d_n(\bm{\lambda})^2$ \eqref{dn2}.
Note the positivity of the quantities,
$\hat{f}_{\ell,n}(\bm{\lambda}),\hat{b}_{\ell,n}(\bm{\lambda}),
s_{\ell}(\bm{\lambda})>0$.

%%%%%%%%%%%%%%%%%%%%%%%%%%%%%%%%%%%%%%%%%%%%%%%%%%%%%%%%%%%%%%%
%                                                             %
%  5. Intertwining Relations                                  %
%                                                             %
%%%%%%%%%%%%%%%%%%%%%%%%%%%%%%%%%%%%%%%%%%%%%%%%%%%%%%%%%%%%%%%
\section{Intertwining Relations}
\label{sec:intertwine}
\setcounter{equation}{0}

Here we demonstrate that the Hamiltonian systems of the original polynomials
reviewed in \S\,\ref{sec:original} and the deformation summarised in
\S\,\ref{sec:deformed} are intertwined by a discrete version of the
Darboux-Crum transformation.
This provides simple expressions of the eigenfunctions of the deformed
systems \eqref{Pln} in terms of those of the original system,
which is exactly solvable.
It also delivers a simple proof of the shape invariance of the deformed
system. The line of arguments goes parallel with those for the other
exceptional orthogonal polynomials \cite{stz,os20}.

First let us discuss the general scheme.
For an adjoint pair of well-defined operators
$\hat{\mathcal{A}}_{\ell}(\bm{\lambda})$ and
$\hat{\mathcal{A}}_{\ell}(\bm{\lambda})^{\dagger}$, let us define
a pair of Hamiltonians $\hat{\mathcal{H}}_{\ell}^{(\pm)}(\bm{\lambda})$
\begin{equation}
  \hat{\mathcal{H}}_{\ell}^{(+)}(\bm{\lambda})\eqdef
  \hat{\mathcal{A}}_{\ell}(\bm{\lambda})^{\dagger}
  \hat{\mathcal{A}}_{\ell}(\bm{\lambda}),\quad
  \hat{\mathcal{H}}_{\ell}^{(-)}(\bm{\lambda})\eqdef
  \hat{\mathcal{A}}_{\ell}(\bm{\lambda})
  \hat{\mathcal{A}}_{\ell}(\bm{\lambda})^{\dagger},
  \label{H+-}
\end{equation}
and consider their Schr\"{o}dinger equations, that is, the eigenvalue
problems:
\begin{equation}
  \hat{\mathcal{H}}_{\ell}^{(\pm)}(\bm{\lambda})
  \hat{\phi}_{\ell,n}^{(\pm)}(x;\bm{\lambda})
  =\hat{\mathcal{E}}_{\ell,n}^{(\pm)}(\bm{\lambda})
  \hat{\phi}_{\ell,n}^{(\pm)}(x;\bm{\lambda})\quad
  (n=0,1,2,\ldots).
  \label{H+-Scheq}
\end{equation}
Obviously the pair of Hamiltonians are intertwined:
\begin{align}
  &\hat{\mathcal{H}}_{\ell}^{(+)}(\bm{\lambda})
  \hat{\mathcal{A}}_{\ell}(\bm{\lambda})^{\dagger}
  =\hat{\mathcal{A}}_{\ell}(\bm{\lambda})^{\dagger}
  \hat{\mathcal{A}}_{\ell}(\bm{\lambda})
  \hat{\mathcal{A}}_{\ell}(\bm{\lambda})^{\dagger}
  =\hat{\mathcal{A}}_{\ell}(\bm{\lambda})^{\dagger}
  \hat{\mathcal{H}}_{\ell}^{(-)}(\bm{\lambda}),\\
  &\hat{\mathcal{A}}_{\ell}(\bm{\lambda})
  \hat{\mathcal{H}}_{\ell}^{(+)}(\bm{\lambda})
  =\hat{\mathcal{A}}_{\ell}(\bm{\lambda})
  \hat{\mathcal{A}}_{\ell}(\bm{\lambda})^{\dagger}
  \hat{\mathcal{A}}_{\ell}(\bm{\lambda})
  =\hat{\mathcal{H}}_{\ell}^{(-)}(\bm{\lambda})
  \hat{\mathcal{A}}_{\ell}(\bm{\lambda}).
\end{align}
If $\hat{\mathcal{A}}_{\ell}(\bm{\lambda})
\hat{\phi}_{\ell,n}^{(+)}(x;\bm{\lambda})\neq 0$ and
$\hat{\mathcal{A}}_{\ell}(\bm{\lambda})^{\dagger}
\hat{\phi}_{\ell,n}^{(-)}(x;\bm{\lambda})\neq 0$, then the two systems are
exactly iso-spectral and there is one-to-one correspondence between the
eigenfunctions:
\begin{align}
  &\hat{\mathcal{E}}_{\ell,n}^{(+)}(\bm{\lambda})
  =\hat{\mathcal{E}}_{\ell,n}^{(-)}(\bm{\lambda}),\\
  &\hat{\phi}_{\ell,n}^{(-)}(x;\bm{\lambda})\propto
  \hat{\mathcal{A}}_{\ell}(\bm{\lambda})
  \hat{\phi}_{\ell,n}^{(+)}(x;\bm{\lambda}),\quad
  \hat{\phi}_{\ell,n}^{(+)}(x;\bm{\lambda})\propto
  \hat{\mathcal{A}}_{\ell}(\bm{\lambda})^{\dagger}
  \hat{\phi}_{\ell,n}^{(-)}(x;\bm{\lambda}).
\end{align}

\bigskip

In the following we will present the explicit forms of the operators
$\hat{\mathcal{A}}_{\ell}(\bm{\lambda})$ and
$\hat{\mathcal{A}}_{\ell}(\bm{\lambda})^{\dagger}$,
which intertwine the original systems in \S\,\ref{sec:original} and
the deformed systems in \S\,\ref{sec:deformed}.
The operators $\hat{\mathcal{A}}_{\ell}(\bm{\lambda})
=(\hat{\mathcal{A}}_{\ell;x,y}(\bm{\lambda}))$
and $\hat{\mathcal{A}}_{\ell}(\bm{\lambda})^{\dagger}
=((\hat{\mathcal{A}}_{\ell}(\bm{\lambda})^{\dagger})_{x,y})
=(\hat{\mathcal{A}}_{\ell;y,x}(\bm{\lambda}))$
($x,y=0,1,\ldots,x_{\text{max}}^{\ell}$) are defined by
\begin{equation}
  \hat{\mathcal{A}}_{\ell}(\bm{\lambda})
  \eqdef\sqrt{\hat{B}_{\ell}(x;\bm{\lambda})}
  -e^{\partial}\sqrt{\hat{D}_{\ell}(x;\bm{\lambda})},\quad
  \hat{\mathcal{A}}_{\ell}(\bm{\lambda})^{\dagger}
  =\sqrt{\hat{B}_{\ell}(x;\bm{\lambda})}
  -\sqrt{\hat{D}_{\ell}(x;\bm{\lambda})}\,e^{-\partial},
  \label{hAl,hAld}
\end{equation}
where $\hat{B}_{\ell}(x;\bm{\lambda})$ and $\hat{D}_{\ell}(x;\bm{\lambda})$
are given by
\begin{align}
  \hat{B}_{\ell}(x;\bm{\lambda})
  &\eqdef B\bigl(x;\mathfrak{t}(\bm{\lambda}+(\ell-1)\bm{\delta})\bigr)
  \frac{\check{\xi}_{\ell}(x+1;\bm{\lambda})}
  {\check{\xi}_{\ell}(x;\bm{\lambda})},
  \label{hBl}\\
  \hat{D}_{\ell}(x;\bm{\lambda})
  &\eqdef D\bigl(x;\mathfrak{t}(\bm{\lambda}+(\ell-1)\bm{\delta})\bigr)
  \frac{\check{\xi}_{\ell}(x-1;\bm{\lambda})}
  {\check{\xi}_{\ell}(x;\bm{\lambda})}.
  \label{hDl}
\end{align}
Compare with the similar expressions for the $X_\ell$ Laguerre \& Jacobi
polynomials (2.10)--(2.15) \& (3.13)--(3.18) in \cite{stz},
and for the $X_\ell$ Wilson \& Askey-Wilson polynomials (3.7)--(3.9) in
\cite{os20}.

Since $\det\hat{\mathcal{A}}_{\ell}(\bm{\lambda})
=\prod_{x=0}^{x_{\text{max}}^{\ell}}\sqrt{\hat{B}_{\ell}(x;\bm{\lambda})}
\neq 0$ for the parameter range under consideration,
the operators $\hat{\mathcal{A}}_{\ell}(\bm{\lambda})$ and
$\hat{\mathcal{A}}_{\ell}(\bm{\lambda})^{\dagger}$ have no zero modes.
By using the two formulas \eqref{xil(l+d)}--\eqref{xil(l)}, we can
show that
\begin{align}
  \hat{\mathcal{H}}_{\ell}^{(+)}(\bm{\lambda})
  &=\hat{\kappa}_{\ell}(\bm{\lambda})
  \bigl(\mathcal{H}(\bm{\lambda}+\ell\bm{\delta}+\bm{\tilde{\delta}})
  +\hat{f}_{\ell,0}(\bm{\lambda})\hat{b}_{\ell,0}(\bm{\lambda})\bigr),
  \label{Hl+=H}\\
  \hat{\mathcal{H}}_{\ell}^{(-)}(\bm{\lambda})
  &=\hat{\kappa}_{\ell}(\bm{\lambda})
  \bigl(\mathcal{H}_{\ell}(\bm{\lambda})
  +\hat{f}_{\ell,0}(\bm{\lambda})\hat{b}_{\ell,0}(\bm{\lambda})\bigr),
  \label{Hl-=Hl}
\end{align}
where $\hat{\kappa}_{\ell}(\bm{\lambda})$ is
\begin{equation}
  \hat{\kappa}_{\ell}(\bm{\lambda})\eqdef
  \left\{
  \begin{array}{ll}
  1&:\text{R}\\
  (abd^{-1}q^{\ell})^{-1}&:\text{$q$R}
  \end{array}\right.\!.
  \label{kappah}
\end{equation}
Therefore the original system with the shifted parameters
($\mathcal{H}(\bm{\lambda}+\ell\bm{\delta}+\tilde{\bm{\delta}})$)
and the deformed system ($\mathcal{H}_{\ell}(\bm{\lambda})$) are
exactly isospectral.
Note that the maximal value of $x$ for
$\mathcal{H}(\bm{\lambda}+\ell\bm{\delta}+\tilde{\bm{\delta}})$
is $N-\ell\, (=x_{\text{max}}^{\ell})$.
Based on the results \eqref{Hl+=H}--\eqref{Hl-=Hl}, we have
\begin{gather}
  \hat{\phi}_{\ell,n}^{(+)}(x;\bm{\lambda})
  =\phi_n(x;\bm{\lambda}+\ell\bm{\delta}+\bm{\tilde{\delta}}),\quad
  \hat{\phi}_{\ell,n}^{(-)}(x;\bm{\lambda})
  =\phi_{\ell,n}(x;\bm{\lambda}),
  \label{phi+-=..}\\
  \hat{\mathcal{E}}_{\ell,n}^{(\pm)}(\bm{\lambda})
  =\hat{\kappa}_{\ell}(\bm{\lambda})\bigl(
  \mathcal{E}_n(\bm{\lambda}+\ell\bm{\delta}+\bm{\tilde{\delta}})
  +\hat{f}_{\ell,0}(\bm{\lambda})\hat{b}_{\ell,0}(\bm{\lambda})\bigr)
  =\hat{\kappa}_{\ell}(\bm{\lambda})\bigl(
  \mathcal{E}_{\ell,n}(\bm{\lambda})
  +\hat{f}_{\ell,0}(\bm{\lambda})\hat{b}_{\ell,0}(\bm{\lambda})\bigr).
  \label{E+-=..}
\end{gather}
The correspondence of the pair of eigenfunctions
$\hat{\phi}_{\ell,n}^{(\pm)}(x)$ with their own normalisation specified
in the preceding sections are related by
\begin{align}
  \hat{\phi}_{\ell,n}^{(-)}(x;\bm{\lambda})
  &=\sqrt{\check{\xi}_{\ell}(1;\bm{\lambda})s_{\ell}(\bm{\lambda})}\,
  \frac{\hat{\mathcal{A}}_{\ell}(\bm{\lambda})
  \hat{\phi}_{\ell,n}^{(+)}(x;\bm{\lambda})}
  {\sqrt{\hat{\kappa}_{\ell}(\bm{\lambda})}\,\hat{f}_{\ell,n}(\bm{\lambda})},
  \\
  \hat{\phi}_{\ell,n}^{(+)}(x;\bm{\lambda})
  &=\frac{1}{\sqrt{\check{\xi}_{\ell}(1;\bm{\lambda})s_{\ell}(\bm{\lambda})}}\,
  \frac{\hat{\mathcal{A}}_{\ell}(\bm{\lambda})^{\dagger}
  \hat{\phi}_{\ell,n}^{(-)}(x;\bm{\lambda})}
  {\sqrt{\hat{\kappa}_{\ell}(\bm{\lambda})}\,\hat{b}_{\ell,n}(\bm{\lambda})}.
  \label{phi+<->phi-}
\end{align}
Let us introduce the operators $\hat{\mathcal{F}}_{\ell}(\bm{\lambda})
=(\hat{\mathcal{F}}_{\ell;x,y}(\bm{\lambda}))$ and
$\hat{\mathcal{B}}_{\ell}(\bm{\lambda})
=(\hat{\mathcal{B}}_{\ell;x,y}(\bm{\lambda}))$
%($x,y=0,1,\ldots,x_{\text{max}}^{\ell}$) defined by
($x,y=0,1,\ldots,$ $x_{\text{max}}^{\ell}$) defined by
\begin{align}
  \hat{\mathcal{F}}_{\ell}(\bm{\lambda})&\eqdef
  \sqrt{\check{\xi}_{\ell}(1;\bm{\lambda})s_{\ell}(\bm{\lambda})}\,
  \psi_{\ell}(x;\bm{\lambda})^{-1}\circ
  \frac{\hat{\mathcal{A}}_{\ell}(\bm{\lambda})}
  {\sqrt{\hat{\kappa}_{\ell}(\bm{\lambda})}}\circ
  \phi_0(x;\bm{\lambda}+\ell\bm{\delta}+\bm{\tilde{\delta}}),
  \label{hatFldef}\\
  \hat{\mathcal{B}}_{\ell}(\bm{\lambda})&\eqdef
  \frac{1}{\sqrt{\check{\xi}_{\ell}(1;\bm{\lambda})s_{\ell}(\bm{\lambda})}}\,
  \phi_0(x;\bm{\lambda}+\ell\bm{\delta}+\bm{\tilde{\delta}})^{-1}\circ
  \frac{\hat{\mathcal{A}}_{\ell}(\bm{\lambda})^{\dagger}}
  {\sqrt{\hat{\kappa}_{\ell}(\bm{\lambda})}}\circ
  \psi_{\ell}(x;\bm{\lambda}).
  \label{hatBldef}
\end{align}
Their explicit forms are:
\begin{align}
  \hat{\mathcal{F}}_{\ell}(\bm{\lambda})&=
  \frac{1}{\varphi(x;\bm{\lambda}+\ell\bm{\delta}+\tilde{\bm{\delta}})}
  \Bigl(v_1^B(x;\bm{\lambda}+\ell\bm{\delta})
  \check{\xi}_{\ell}(x;\bm{\lambda})e^{\partial}
  -v_1^D(x;\bm{\lambda}+\ell\bm{\delta})
  \check{\xi}_{\ell}(x+1;\bm{\lambda})\Bigr),
  \label{hatFlform}\\
  \hat{\mathcal{B}}_{\ell}(\bm{\lambda})&=
  \frac{1}{\check{\xi}_{\ell}(x;\bm{\lambda})}
  \frac{1}{\varphi(x;\bm{\lambda}+(\ell-1)\bm{\lambda}+\tilde{\bm{\delta}})}
  \Bigl(v_2^B(x;\bm{\lambda}+(\ell-1)\bm{\delta})
  -v_2^D(x;\bm{\lambda}+(\ell-1)\bm{\delta})e^{-\partial}\Bigr).
 \label{hatBlform}
\end{align}
Compare with the similar expressions
for the $X_\ell$ Wilson \& Askey-Wilson polynomials (3.20)--(3.21) in
\cite{os20}.
The operators $\hat{\mathcal{F}}_{\ell}(\bm{\lambda})$ and
$\hat{\mathcal{B}}_{\ell}(\bm{\lambda})$ act as the forward and backward
shift operators connecting the original polynomials $P_n$ and
the exceptional polynomials $P_{\ell,n}$:
\begin{align}
  \hat{\mathcal{F}}_{\ell}(\bm{\lambda})
  \check{P}_n(x;\bm{\lambda}+\ell\bm{\delta}+\bm{\tilde{\delta}})
  &=\hat{f}_{\ell,n}(\bm{\lambda})\check{P}_{\ell,n}(x;\bm{\lambda}),
  \label{FhatPn=Pln}\\
  \hat{\mathcal{B}}_{\ell}(\bm{\lambda})\check{P}_{\ell,n}(x;\bm{\lambda})
  &=\hat{b}_{\ell,n}(\bm{\lambda})
  \check{P}_n(x;\bm{\lambda}+\ell\bm{\delta}+\bm{\tilde{\delta}}).
  \label{BhatPln=Pn}
\end{align}
The former relation \eqref{FhatPn=Pln} with the explicit form of
$\hat{\mathcal{F}}_{\ell}(\bm{\lambda})$ \eqref{hatFlform} provides the
explicit expression \eqref{Pln} of the exceptional orthogonal polynomials.
In terms of $\hat{\mathcal{F}}_{\ell}(\bm{\lambda})$ and
$\hat{\mathcal{B}}_{\ell}(\bm{\lambda})$,
the relations \eqref{Hl+=H}--\eqref{Hl-=Hl} become
\begin{align}
  &\hat{\mathcal{B}}_{\ell}(\bm{\lambda})
  \hat{\mathcal{F}}_{\ell}(\bm{\lambda})
  =\widetilde{\mathcal{H}}(\bm{\lambda}+\ell\bm{\delta}+\tilde{\bm{\delta}})
  +\hat{f}_{\ell,0}(\bm{\lambda})\hat{b}_{\ell,0}(\bm{\lambda}),\\
  &\hat{\mathcal{F}}_{\ell}(\bm{\lambda})
  \hat{\mathcal{B}}_{\ell}(\bm{\lambda})
  =\widetilde{\mathcal{H}}_{\ell}(\bm{\lambda})
  +\hat{f}_{\ell,0}(\bm{\lambda})\hat{b}_{\ell,0}(\bm{\lambda}).
\end{align}
The other simple consequences of these relations are
\begin{equation}
  \hat{\mathcal{E}}_{\ell,n}^{(\pm)}(\bm{\lambda})
  =\hat{\kappa}_{\ell}(\bm{\lambda})\hat{f}_{\ell,n}(\bm{\lambda})
  \hat{b}_{\ell,n}(\bm{\lambda}),\quad
   \mathcal{E}_n(\bm{\lambda}+\ell\bm{\delta})
  =\hat{f}_{\ell,n}(\bm{\lambda})\hat{b}_{\ell,n}(\bm{\lambda})
  -\hat{f}_{\ell,0}(\bm{\lambda})\hat{b}_{\ell,0}(\bm{\lambda}).
  \label{Eln+-}
\end{equation}

The $\ell^2$ inner product for $\phi_{\ell,n}$ and $\phi_{\ell,m}$
can be calculated in the following way:
\begin{align}
  &\quad\bigl(\phi_{\ell,n}(\,\cdot\,;\bm{\lambda}),
  \phi_{\ell,m}(\,\cdot\,;\bm{\lambda})\bigr)\n
  &=\frac{1}{\hat{f}_{\ell,m}(\bm{\lambda})}
  \sqrt{\frac{\check{\xi}_{\ell}(1;\bm{\lambda})s_{\ell}(\bm{\lambda})}
  {\hat{\kappa}_{\ell}(\bm{\lambda})}}
  \bigl(\phi_{\ell,n}(\,\cdot\,;\bm{\lambda}),
  \hat{\mathcal{A}}_{\ell}(\bm{\lambda})
  \phi_m(\,\cdot\,;\bm{\lambda}+\ell\bm{\delta}+\tilde{\bm{\delta}})\bigr)\n
  &=\frac{1}{\hat{f}_{\ell,m}(\bm{\lambda})}
  \sqrt{\frac{\check{\xi}_{\ell}(1;\bm{\lambda})s_{\ell}(\bm{\lambda})}
  {\hat{\kappa}_{\ell}(\bm{\lambda})}}
  \bigl(\hat{\mathcal{A}}_{\ell}(\bm{\lambda})^{\dagger}
  \phi_{\ell,n}(\,\cdot\,;\bm{\lambda}),
  \phi_m(\,\cdot\,;\bm{\lambda}+\ell\bm{\delta}+\tilde{\bm{\delta}})\bigr)\n
  &=\frac{\hat{b}_{\ell,n}(\bm{\lambda})}{\hat{f}_{\ell,m}(\bm{\lambda})}
  \check{\xi}_{\ell}(1;\bm{\lambda})s_{\ell}(\bm{\lambda})
  \bigl(\phi_n(\,\cdot\,;\bm{\lambda}+\ell\bm{\delta}+\tilde{\bm{\delta}}),
  \phi_m(\,\cdot\,;\bm{\lambda}+\ell\bm{\delta}+\tilde{\bm{\delta}})\bigr)\n
  &=\check{\xi}_{\ell}(1;\bm{\lambda})\,
  \frac{\delta_{nm}}{d_n(\bm{\lambda}+\ell\bm{\delta}+\tilde{\bm{\delta}})^2}
  \frac{\hat{b}_{\ell,n}(\bm{\lambda})}{\hat{f}_{\ell,n}(\bm{\lambda})}
  s_{\ell}(\bm{\lambda}),
\end{align}
where we have used \eqref{phi+-=..}, \eqref{phi+<->phi-} and \eqref{ortho}.
This gives a proof of \eqref{dln2}.

It is interesting to note that the operator
$\hat{\mathcal{A}}_{\ell}(\bm{\lambda})$ intertwines those of the original
and deformed systems $\mathcal{A}(\bm{\lambda})$ and
$\mathcal{A}_{\ell}(\bm{\lambda})$:
\begin{align}
  &\hat{\mathcal{A}}_{\ell}(\bm{\lambda}+\bm{\delta})
  \mathcal{A}(\bm{\lambda}+\ell\bm{\delta}+\bm{\tilde{\delta}})
  =\mathcal{A}_{\ell}(\bm{\lambda})
  \hat{\mathcal{A}}_{\ell}(\bm{\lambda}),
  \label{AhA=AlAh}\\
  &\hat{\mathcal{A}}_{\ell}(\bm{\lambda})
  \mathcal{A}(\bm{\lambda}+\ell\bm{\delta}+\bm{\tilde{\delta}})^{\dagger}
  =\mathcal{A}_{\ell}(\bm{\lambda})^{\dagger}
  \hat{\mathcal{A}}_{\ell}(\bm{\lambda}+\bm{\delta}).
  \label{AhAd=AldAh}
\end{align}
In terms of the definitions of the forward shift operators
$\mathcal{F}(\bm{\lambda})$ \eqref{Fdef},
$\mathcal{F}_{\ell}(\bm{\lambda})$ \eqref{Fldef},
$\hat{\mathcal{F}}_{\ell}(\bm{\lambda})$ \eqref{hatFldef}, and
$\mathcal{B}(\bm{\lambda})$ \eqref{Bdef},
$\mathcal{B}_{\ell}(\bm{\lambda})$ \eqref{Bldef},
the above relations are rewritten as:
\begin{align}
  &\hat{s}_{\ell}(\bm{\lambda}+\bm{\delta})
  \hat{\mathcal{F}}_{\ell}(\bm{\lambda}+\bm{\delta})
  \mathcal{F}(\bm{\lambda}+\ell\bm{\delta}+\bm{\tilde{\delta}})
  =\hat{s}_{\ell}(\bm{\lambda})
  \mathcal{F}_{\ell}(\bm{\lambda})
  \hat{\mathcal{F}}_{\ell}(\bm{\lambda}),
  \label{FlhF=FlFlh}\\
  &\hat{s}_{\ell}(\bm{\lambda})
  \hat{\mathcal{F}}_{\ell}(\bm{\lambda})
  \mathcal{B}(\bm{\lambda}+\ell\bm{\delta}+\bm{\tilde{\delta}})
  =\hat{s}_{\ell}(\bm{\lambda}+\bm{\delta})
  \mathcal{B}_{\ell}(\bm{\lambda})
  \hat{\mathcal{F}}_{\ell}(\bm{\lambda}+\bm{\delta}),
  \label{FlhB=BlFlh}
\end{align}
where $\hat{s}_{\ell}(\bm{\lambda})$ is
\begin{equation}
  \hat{s}_{\ell}(\bm{\lambda})
  \eqdef\hat{\kappa}_{\ell}(\bm{\lambda})\times\left\{
  \begin{array}{ll}
  c+\ell-1&:\text{R}\\
  1-cq^{\ell-1}&:\text{$q$R}
  \end{array}\right.\!.
\end{equation}
These relations can be proven by explicit calculation with the help of
the two formulas of the deforming polynomial
$\check{\xi}_{\ell}(x;\bm{\lambda})$ \eqref{xil(l+d)}--\eqref{xil(l)}.

By applying $\hat{\mathcal{A}}_{\ell}(\bm{\lambda}+\bm{\delta})$
and $\hat{\mathcal{A}}_{\ell}(\bm{\lambda})$ to \eqref{Aphi=fphi}
and \eqref{Adphi=bphi} (with replacement
$\bm{\lambda}\to\bm{\lambda}+\ell\bm{\delta}+\tilde{\bm{\delta}}$)
respectively, together with the use of \eqref{AhA=AlAh},
\eqref{AhAd=AldAh} and \eqref{phi+<->phi-}, we obtain
\begin{align}
  \mathcal{A}_{\ell}(\bm{\lambda})\phi_{\ell,n}(x;\bm{\lambda})
  &=\sqrt{\frac{\hat{\kappa}_{\ell}(\bm{\lambda}+\bm{\delta})}
  {\hat{\kappa}_{\ell}(\bm{\lambda})}
  \frac{s_{\ell}(\bm{\lambda})}{s_{\ell}(\bm{\lambda}+\bm{\delta})}
  \frac{\check{\xi}_{\ell}(1;\bm{\lambda})}
  {\check{\xi}_{\ell}(1;\bm{\lambda}+\bm{\delta})}
  \frac{1}{B(0;\bm{\lambda}+\ell\bm{\delta}+\tilde{\bm{\delta}})}}\,
  \frac{\hat{f}_{\ell,n-1}(\bm{\lambda}+\bm{\delta})}
  {\hat{f}_{\ell,n}(\bm{\lambda})}\n
  &\qquad\qquad\times
  f_n(\bm{\lambda}+\ell\bm{\delta}+\bm{\tilde{\delta}})
  \phi_{\ell,n-1}(x;\bm{\lambda}+\bm{\delta})\n
  &=\frac{1}{\sqrt{B_{\ell}(0;\bm{\lambda})}}\,
  f_n(\bm{\lambda}+\ell\bm{\delta})
  \phi_{\ell,n-1}(x;\bm{\lambda}+\bm{\delta}),
  \label{Alphiln=fnphiln2}\\
  \mathcal{A}_{\ell}(\bm{\lambda})^{\dagger}
  \phi_{\ell,n-1}(x;\bm{\lambda}+\bm{\delta})
  &=\sqrt{\frac{\hat{\kappa}_{\ell}(\bm{\lambda})}
  {\hat{\kappa}_{\ell}(\bm{\lambda}+\bm{\delta})}
  \frac{s_{\ell}(\bm{\lambda}+\bm{\delta})}{s_{\ell}(\bm{\lambda})}
  \frac{\check{\xi}_{\ell}(1;\bm{\lambda}+\bm{\delta})}
  {\check{\xi}_{\ell}(1;\bm{\lambda})}
  B(0;\bm{\lambda}+\ell\bm{\delta}+\tilde{\bm{\delta}})}\,
  \frac{\hat{f}_{\ell,n}(\bm{\lambda})}
  {\hat{f}_{\ell,n-1}(\bm{\lambda}+\bm{\delta})}\n
  &\qquad\qquad\times
  b_{n-1}(\bm{\lambda}+\ell\bm{\delta}+\bm{\tilde{\delta}})
  \phi_{\ell,n}(x;\bm{\lambda})\n
  &=\sqrt{B_{\ell}(0;\bm{\lambda})}\,
  b_{n-1}(\bm{\lambda}+\ell\bm{\delta})
  \phi_{\ell,n}(x;\bm{\lambda}+\bm{\delta}).
  \label{Aldphiln=bnphiln2}
\end{align}
In the calculation use is made of the explicit forms of
$\hat{\kappa}_{\ell}(\bm{\lambda})$, $s_{\ell}(\bm{\lambda})$,
$B_{\ell}(x;\bm{\lambda})$, $\hat{f}_{\ell,n}(\bm{\lambda})$,
$f_n(\bm{\lambda})$ and $b_n(\bm{\lambda})$ in the second equalities.
This provides a proof of \eqref{Alphiln=flnphiln}--\eqref{fln,bln}
without recourse to the shape invariance of the deformed system.
Likewise the above intertwining relations of the forward-backward
shift operators \eqref{FlhF=FlFlh}--\eqref{FlhB=BlFlh} give a proof of
\eqref{FlPln=flnPln}--\eqref{BlPln=blnPln}, respectively, again
without recourse to the shape invariance.

Since the $q$-Racah polynomial $\check{P}^{\text{$q$R}}_n(x;\bm{\lambda})$
\eqref{Pn=R,qR} is related to the Askey-Wilson polynomial
$p_n(\cos x;a,b,c,d|q)$ as \cite{koeswart}
\begin{equation}
  \check{P}^{\text{$q$R}}_n(x;\bm{\lambda})
  =\frac{d^{\frac{n}{2}}}{(a,b,c\,;q)_n}\,
  p_n\bigl(\tfrac12(d^{\frac12}q^x+d^{-\frac12}q^{-x});
  ad^{-\frac12},bd^{-\frac12},cd^{-\frac12},d^{\frac12}|\,q\bigr),
\end{equation}
many formulas for the $q$-Racah case in sections \ref{sec:deformed} and
\ref{sec:intertwine} are obtained essentially from those for the
Askey-Wilson case \cite{os20} by the following replacement:
\begin{equation}
  e^{ix^{\text{AW}}}=d^{\frac12}q^{x+\frac12\ell},\quad
  q^{\bm{\lambda}^{\text{AW}}}
  =(ad^{-\frac12},bd^{-\frac12},cd^{-\frac12},d^{\frac12}).
\end{equation}

%%%%%%%%%%%%%%%%%%%%%%%%%%%%%%%%%%%%%%%%%%%%%%%%%%%%%%%%%%%%%%%%%
%                                                               %
%  6. Other $X_{\ell}$ Polynomials:                             %
%     dual ($q$-)Hahn, little $q$-Jacobi                        %
%                                                               %
%%%%%%%%%%%%%%%%%%%%%%%%%%%%%%%%%%%%%%%%%%%%%%%%%%%%%%%%%%%%%%%%%
\section{Other $X_{\ell}$ Polynomials: dual ($q$)-Hahn, little $q$-Jacobi}
\label{sec:other}
\setcounter{equation}{0}

In \S\,\ref{sec:original}--\S\,\ref{sec:intertwine} we have derived the
exceptional Racah and $q$-Racah Hamiltonian systems by deforming
those of the Racah and $q$-Racah in parallel in terms of  a degree $\ell$
polynomial with twisted parameters.
It is well known that the Racah polynomials can be obtained from the
$q$-Racah polynomials by taking the standard $q\to 1$ limit with an
appropriate overall rescaling. The same limiting procedure could be applied
to derive the exceptional Racah polynomials from the exceptional $q$-Racah
polynomials.

Likewise various orthogonal polynomials of a discrete variable can be
obtained from the $q$-Racah polynomials by many different limiting
procedures with/without the $q\to 1$ limit.
Here we present two such examples: the dual ($q$)-Hahn and the little
$q$-Jacobi polynomials and the corresponding exceptional polynomials.
The former is a finite dimensional example and the latter is infinite
dimensional.
It should be stressed, however, that there is no guarantee that the
limiting procedure among the undeformed polynomials could be lifted to
produce the corresponding exceptional polynomials.
For example, the Hermite polynomials are known to be obtained from the
Jacobi or the Laguerre polynomials by a certain limit procedure.
But that does not produce exceptional Hermite polynomials from the known
exceptional Jacobi or Laguerre polynomials.

%%%%%%%%%%%%%%%%%%%%%%%%%%%%%%%%%%%%%%%%%%%
%                                         %
% 6.1 dual ($q$-)Hahn                     %
%                                         %
%%%%%%%%%%%%%%%%%%%%%%%%%%%%%%%%%%%%%%%%%%%
\subsection{Dual ($q$)-Hahn}
\label{sec:dqH}

In this subsection we present the ordinary and the exceptional dual Hahn
(dH) and the dual $q$-Hahn (d$q$H) polynomials.
Like as ($q$)-Racah cases, these are finite dimensional:
$x_{\text{max}}=n_{\text{max}}=N$ and
$x_{\text{max}}^{\ell}=n_{\text{max}}^{\ell}=N-\ell$.
The dual $q$-Hahn case is obtained from the $q$-Racah case by the
following limit:
\begin{equation}
  q^{\bm{\lambda}^{\text{$q$R}}}=(q^{-N},a,t,abq^{-1}),\quad
  \text{$q$R}\ \xrightarrow{t\to 0}\ \text{d$q$H}.
\end{equation}
The dual Hahn case is obtained from the dual $q$-Hahn case
by taking $q\to 1$ limit with an appropriate overall rescaling.

%%%%%%%%%%%%%%%%%%%%%%%%%%%%%%%%%%%%%%%%%%%
% 6.1.1 original system: dual ($q$-)Hahn  %
%%%%%%%%%%%%%%%%%%%%%%%%%%%%%%%%%%%%%%%%%%%
\subsubsection{Original systems}

The Hamiltonian systems thus obtained belong to the $\epsilon=1$ case
of \cite{os12} and they are listed  as follows:
\begin{align}
  &\left\{
  \begin{array}{ll}
  \bm{\lambda}=(a,b,N)&:\text{dH}\\
  q^{\bm{\lambda}}=(a,b,q^N)&:\text{d$q$H}
  \end{array}\right.\!,\quad
  \bm{\delta}=(1,0,-1)\ :\text{dH,\,d$q$H},\quad
  \kappa=\left\{
  \begin{array}{ll}
  1&:\text{dH}\\
  q^{-1}&:\text{d$q$H}
  \end{array}\right.\!,\\
  &\left\{
  \begin{array}{ll}
  a>0,\ b>0&:\text{dH}\\
  0<a<1,\ 0<b<1,&:\text{d$q$H}
  \end{array}\right.\!,
  \label{dqH:parameter}\\
  &B(x;\bm{\lambda})=\left\{
  \begin{array}{ll}
  {\displaystyle
  \frac{(x+a)(x+a+b-1)(N-x)}{(2x-1+a+b)(2x+a+b)}}&:\text{dH}\\[8pt]
  {\displaystyle
  \frac{(q^{x-N}-1)(1-aq^x)(1-abq^{x-1})}
  {(1-abq^{2x-1})(1-abq^{2x})}}&:\text{d$q$H}
  \end{array}\right.\!,\\
  &D(x;\bm{\lambda})=\left\{
  \begin{array}{ll}
  {\displaystyle
  \frac{x(x+b-1)(x+a+b+N-1)}{(2x-2+a+b)(2x-1+a+b)}}&:\text{dH}\\[8pt]
  {\displaystyle
  aq^{x-N-1}\frac{(1-q^x)(1-abq^{x+N-1})(1-bq^{x-1})}
  {(1-abq^{2x-2})(1-abq^{2x-1})}}&:\text{d$q$H}
  \end{array}\right.\!,\\
  &\mathcal{E}_n(\bm{\lambda})=\left\{
  \begin{array}{ll}
  n&:\text{dH}\\
  q^{-n}-1&:\text{d$q$H}
  \end{array}\right.\!,\quad
  \eta(x;\bm{\lambda})=\left\{
  \begin{array}{ll}
  x(x+a+b-1)&:\text{dH}\\
  (q^{-x}-1)(1-abq^{x-1})&:\text{d$q$H}
  \end{array}\right.\!,\\
  &\check{P}_n(x;\bm{\lambda})
  =P_n(\eta(x;\bm{\lambda});\bm{\lambda})=\left\{
  \begin{array}{ll}
  {\displaystyle
  {}_3F_2\Bigl(
  \genfrac{}{}{0pt}{}{-n,\,x+a+b-1,\,-x}
  {a,\,-N}\Bigm|1\Bigr)}&:\text{dH}\\[8pt]
  {\displaystyle
  {}_3\phi_2\Bigl(
  \genfrac{}{}{0pt}{}{q^{-n},\,abq^{x-1},\,q^{-x}}
  {a,\,q^{-N}}\Bigm|q\,;q\Bigr)}&:\text{d$q$H}
  \end{array}\right.\n
  &\phantom{\check{P}_n(x;\bm{\lambda})
  =P_n(\eta(x;\bm{\lambda});\bm{\lambda})}=\left\{
  \begin{array}{ll}
  {\displaystyle
  R_n(\eta(x;\bm{\lambda})\,;a-1,b-1,N)}&:\text{dH}\\[4pt]
  {\displaystyle
  R_n(1+abq^{-1}+\eta(x;\bm{\lambda})\,;aq^{-1},bq^{-1},N|q)}&:\text{d$q$H}
  \end{array}\right.\!,\\
  &\phi_0(x;\bm{\lambda})^2=\left\{
  \begin{array}{ll}
  {\displaystyle
  \frac{N!}{x!\,(N-x)!}\,
  \frac{(a)_x\,(2x+a+b-1)(a+b)_N}{(b)_x\,(x+a+b-1)_{N+1}\,}}&:\text{dH}\\[8pt]
  {\displaystyle
  \frac{(q\,;q)_N}{(q\,;q)_x\,(q\,;q)_{N-x}}\,
  \frac{(a,abq^{-1}\,;q)_x}{(abq^N,b\,;q)_x\,a^x}\,
  \frac{1-abq^{2x-1}}{1-abq^{-1}}}&:\text{d$q$H}
  \end{array}\right.\!,\\
  &d_n(\bm{\lambda})^2=\left\{
  \begin{array}{ll}
  {\displaystyle
  \frac{N!}{n!\,(N-n)!}\,\frac{(a)_n\,(b)_{N-n}}{(b)_N}
  \times\frac{(b)_{N}}{(a+b)_N}}&:\text{dH}\\[8pt]
  {\displaystyle
  \frac{(q\,;q)_N}{(q\,;q)_n\,(q\,;q)_{N-n}}\,
  \frac{(a\,;q)_n(b\,;q)_{N-n}}{(b;q)_N\,a^n}
  \times\frac{(b\,;q)_N\,a^N}{(ab;q)_N}}&:\text{d$q$H}
  \end{array}\right.\!,\\
  &\varphi(x;\bm{\lambda})=\left\{
  \begin{array}{ll}
  {\displaystyle
  \frac{2x+a+b}{a+b}}&:\text{dH}\\[6pt]
  {\displaystyle
  \frac{q^{-x}-abq^x}{1-ab}}&:\text{d$q$H}
  \end{array}\right.\!,\quad
  f_n(\bm{\lambda})=\mathcal{E}_n(\bm{\lambda}),
  \ b_n(\bm{\lambda})=1\ :\text{dH,\,d$q$H}.
\end{align}

%%%%%%%%%%%%%%%%%%%%%%%%%%%%%%%%%%%%%%%%%%%
% 6.1.2 deformed system: dual ($q$-)Hahn  %
%%%%%%%%%%%%%%%%%%%%%%%%%%%%%%%%%%%%%%%%%%%
\subsubsection{Deformed systems}

We restrict the parameter range of \eqref{dqH:parameter} as follows:
\begin{equation}
  \left\{
  \begin{array}{ll}
  a>0,\ b>1&:\text{dH}\\
  0<a<1,\ 0<b<q,&:\text{d$q$H}
  \end{array}\right.\!.
\end{equation}
The data for the Hamiltonian systems of the exceptional dual ($q$)-Hahn
polynomials are as follows:
\begin{align}
  &\check{\xi}_{\ell}(x;\bm{\lambda})
  =\xi_{\ell}(\eta(x;\bm{\lambda}+(\ell-1)\bm{\delta});\bm{\lambda})\n
  &\phantom{\check{\xi}_{\ell}(x;\bm{\lambda})}
  =\check{P}_{\ell}\bigl(x;\mathfrak{t}
  \bigl(\bm{\lambda}+(\ell-1)\bm{\delta}\bigr)\bigr),\quad
  \mathfrak{t}(\bm{\lambda})\eqdef
  (\lambda_1+\lambda_2+\lambda_3-1,1-\lambda_3,1-\lambda_2)
  \ \ :\text{dH\,d$q$H}\n
  &\phantom{\check{\xi}_{\ell}(x;\bm{\lambda})}
  =\left\{
  \begin{array}{ll}
  {\displaystyle
  {}_3F_2\Bigl(
  \genfrac{}{}{0pt}{}{-\ell,\,a+b+x+\ell-2,\,-x}
  {a+b+N-1,\,b-1}\Bigm|1\Bigr)}&:\text{dH}\\[8pt]
  {\displaystyle
  {}_3\phi_2\Bigl(
  \genfrac{}{}{0pt}{}{q^{-\ell},\,abq^{x+\ell-2},\,q^{-x}}
  {abq^{N-1},\,bq^{-1}}\Bigm|q\,;q\Bigr)}&:\text{d$q$H}
  \end{array}\right.\!,\\
  &v_1^B(x;\bm{\lambda})=\left\{
  \begin{array}{ll}
  {\displaystyle
  \frac{(x-N)(x+a)}{a+b-1}}&:\text{dH}\\[6pt]
  {\displaystyle
  q^{-x}\frac{(1-q^{x-N})(1-aq^x)}{1-abq^{-1}}}&:\text{d$q$H}
  \end{array}\right.\!,\\
  &v_2^B(x;\bm{\lambda})=\left\{
  \begin{array}{ll}
  {\displaystyle
  \frac{x+a+b-1}{a+b-1}}&:\text{dH}\\[6pt]
  {\displaystyle
  q^{-x}\frac{1-abq^{x-1}}{1-abq^{-1}}}&:\text{d$q$H}
  \end{array}\right.\!,\\
  &v_1^D(x;\bm{\lambda})=\left\{
  \begin{array}{ll}
  {\displaystyle
  \frac{(x+a+b+N-1)(x+b-1)}{a+b-1}}&:\text{dH}\\[6pt]
  {\displaystyle
  q^{-x}b^{-1}q^{1-N}\frac{(1-abq^{x+N-1})(1-bq^{x-1})}{1-abq^{-1}}}
  &:\text{d$q$H}
  \end{array}\right.\!,\\
  &v_2^D(x;\bm{\lambda})=\left\{
  \begin{array}{ll}
  {\displaystyle
  -\frac{x}{a+b-1}}&:\text{dH}\\[6pt]
  {\displaystyle
  -abq^{-1}\frac{1-q^x}{1-abq^{-1}}}&:\text{d$q$H}
  \end{array}\right.\!,\\
  &\tilde{\bm{\delta}}=(0,-1,0)\ :\text{dH,\,d$q$H},\\
  &\hat{f}_{\ell,n}(\bm{\lambda})=\left\{
  \begin{array}{ll}
  {\displaystyle
  -b-N+n+1}&:\text{dH}\\[2pt]
  {\displaystyle
  -b^{-1}q^{1-N}(1-bq^{N-n-1})}&:\text{d$q$H}
  \end{array}\right.\!,\quad
  \hat{b}_{\ell,n}(\bm{\lambda})=1\ :\text{dH,\,d$q$H},\\
  &\hat{\kappa}_{\ell}(\bm{\lambda})=\left\{
  \begin{array}{ll}
  1&:\text{dH}\\
  bq^{N-\ell-1}&:\text{d$q$H}
  \end{array}\right.\!,\quad
  s_{\ell}(\bm{\lambda})=\left\{
  \begin{array}{ll}
  {\displaystyle
  (1-b)\frac{a+b+N-1}{a+b+\ell-1}}&:\text{dH}\\[6pt]
  {\displaystyle
  q^{\ell-N}(1-b^{-1}q)\frac{1-abq^{N-1}}{1-abq^{\ell-1}}}&:\text{d$q$H}
  \end{array}\right.\!,\\
  &\hat{s}_{\ell}(\bm{\lambda})=\hat{\kappa}_{\ell}(\bm{\lambda})
  \ :\text{dH,\,d$q$H}.
\end{align}
Note that $\hat{f}_{\ell,n}(\bm{\lambda}),s_{\ell}(\bm{\lambda})<0$
and $\hat{b}_{\ell,n}(\bm{\lambda})>0$.
All the formulas in \S\,\ref{sec:original}--\S\,\ref{sec:intertwine}
are satisfied.

%%%%%%%%%%%%%%%%%%%%%%%%%%%%%%%%%%%%%%%%%%%
%                                         %
% 6.2 little $q$-Jacobi                   %
%                                         %
%%%%%%%%%%%%%%%%%%%%%%%%%%%%%%%%%%%%%%%%%%%
\subsection{Little $q$-Jacobi}
\label{sec:lqJ}

In this subsection we present the ordinary and the exceptional
little $q$-Jacobi (l$q$J) polynomials.
They are infinite dimensional: $x_{\text{max}}=n_{\text{max}}=\infty$ and
$x_{\text{max}}^{\ell}=n_{\text{max}}^{\ell}=\infty$.
The Hamiltonian system of the  little $q$-Jacobi polynomials is obtained from
that of the $q$-Racah polynomials by the following limit:
\begin{equation}
  q^{\bm{\lambda}^{\text{$q$R}}}=(q^{-N},aq^{N+1}t^{-1},bq,t^{-1}),\quad
  \text{$q$R}\ \xrightarrow{t\to 0}\ \text{al$q$H}
  \ \xrightarrow{N\to\infty}\ \text{l$q$J},
\end{equation}
where al$q$H stands for the alternative $q$-Hahn system
(with $\bm{\lambda}=(aq,bq,N)$) in \S\,5.3.1 of \cite{os12}.

%%%%%%%%%%%%%%%%%%%%%%%%%%%%%%%%%%%%%%%%%%%%
% 6.2.1 original system: little $q$-Jacobi %
%%%%%%%%%%%%%%%%%%%%%%%%%%%%%%%%%%%%%%%%%%%%
\subsubsection{Original system}

The data of the shape invariant Hamiltonian system whose eigenfunctions are
described by the little $q$-Jacobi polynomials are as follows \cite{os12}:
\begin{align}
  &q^{\bm{\lambda}}=(a,b),\quad
  \bm{\delta}=(1,1),\quad \kappa=q^{-1},\quad 0<a<q^{-1},\ \ 0<b<q^{-1},\\
  &B(x;\bm{\lambda})=a(q^{-x}-bq),\quad
  D(x;\bm{\lambda})=q^{-x}-1,\\
  &\mathcal{E}_n(\bm{\lambda})=(q^{-n}-1)(1-abq^{n+1}),\quad
  \eta(x;\bm{\lambda})=1-q^x,\\
  &\check{P}_n(x;\bm{\lambda})=P_n(\eta(x;\bm{\lambda});\bm{\lambda})
  ={}_3\phi_1\Bigl(
  \genfrac{}{}{0pt}{}{q^{-n},abq^{n+1},q^{-x}}{bq}
  \Bigm|q\,;a^{-1}q^x\Bigr)\n
  &\phantom{\check{P}_n(x;\bm{\lambda})=P_n(\eta(x;\bm{\lambda});\bm{\lambda})}
  =(-a)^{-n}q^{-\frac12n(n+1)}\frac{(aq\,;q)_n}{(bq\,;q)_n}\,
  {}_2\phi_1\Bigl(
  \genfrac{}{}{0pt}{}{q^{-n},\,abq^{n+1}}{aq}\Bigm|q\,;q^{x+1}\Bigr)\n
  &\phantom{\check{P}_n(x;\bm{\lambda})=P_n(\eta(x;\bm{\lambda});\bm{\lambda})}
  =(-a)^{-n}q^{-\frac12n(n+1)}\frac{(aq\,;q)_n}{(bq\,;q)_n}\,
  p_n(1-\eta(x;\bm{\lambda});a,b|q),
  \label{littleqjacobinorm}\\
  &\phi_0(x;\bm{\lambda})^2=\frac{(bq\,;q)_x}{(q\,;q)_x}(aq)^x,
  \label{littleqjacobiphi0}\\
  &d_n(\bm{\lambda})^2
  =\frac{(bq,abq\,;q)_n\,a^nq^{n^2}}{(q,aq\,;q)_n}\,
  \frac{1-abq^{2n+1}}{1-abq}
  \times\frac{(aq\,;q)_{\infty}}{(abq^2\,;q)_{\infty}}\,,
  \label{littleqjacobidn}\\
  &\varphi(x;\bm{\lambda})=q^x,\quad
  f_n(\bm{\lambda})=\mathcal{E}_n(\bm{\lambda}),\quad
  b_n(\bm{\lambda})=1.
\end{align}

%%%%%%%%%%%%%%%%%%%%%%%%%%%%%%%%%%%%%%%%%%%%
% 6.2.2 deformed system: little $q$-Jacobi %
%%%%%%%%%%%%%%%%%%%%%%%%%%%%%%%%%%%%%%%%%%%%
\subsubsection{Deformed system}

The data for the exceptional little $q$-Jacobi polynomials are as follows:
\begin{align}
  &\check{\xi}_{\ell}(x;\bm{\lambda})
  =\xi_{\ell}(\eta(x;\bm{\lambda}+(\ell-1)\bm{\delta});\bm{\lambda})\n
  &\phantom{\check{\xi}_{\ell}(x;\bm{\lambda})}
  =\check{P}_{\ell}(x;\mathfrak{t}(\bm{\lambda}+(\ell-1)\bm{\delta})),\quad
  \mathfrak{t}(\bm{\lambda})\eqdef(-\lambda_1-2,\lambda_2)\n
  &\phantom{\check{\xi}_{\ell}(x;\bm{\lambda})}
  ={}_3\phi_1\Bigl(
  \genfrac{}{}{0pt}{}{q^{-\ell},a^{-1}bq^{\ell-1},q^{-x}}{bq^{\ell}}
  \Bigm|q\,;aq^{x+\ell+1}\Bigr),\\
  &v_1^B(x;\bm{\lambda})=-aq^{x+1},\quad
  v_2^B(x;\bm{\lambda})=1-bq^{x+1},\\
  &v_1^D(x;\bm{\lambda})=-q^x,\quad
  v_2^D(x;\bm{\lambda})=1-q^x,\\
  &\tilde{\bm{\delta}}=(1,-1),\\
  &\hat{f}_{\ell,n}(\bm{\lambda})=q^{-n}(1-aq^{n+1})
  \frac{1-bq^{2\ell+n}}{1-bq^{\ell}},\quad
  \hat{b}_{\ell,n}(\bm{\lambda})=1-bq^{\ell},\\
  &\hat{\kappa}_{\ell}(\bm{\lambda})=(aq^{\ell+1})^{-1},\quad
  s_{\ell}(\bm{\lambda})=\frac{1}{1-bq^{\ell}},\\
  &\hat{s}_{\ell}(\bm{\lambda})
  =\hat{\kappa}_{\ell}(\bm{\lambda})(1-bq^{\ell}).
\end{align}
Note that $\hat{f}_{\ell,n}(\bm{\lambda}),\hat{b}_{\ell,n}(\bm{\lambda}),
s_{\ell}(\bm{\lambda})>0$.
All the formulas in \S\,\ref{sec:original}--\S\,\ref{sec:intertwine}
are satisfied.

%%%%%%%%%%%%%%%%%%%%%%%%%%%%%%%%%%%%%%%%%%%%%%%%%%%%%%%%%%%%%%%
%                                                             %
%  7. Summary and Comments                                    %
%                                                             %
%%%%%%%%%%%%%%%%%%%%%%%%%%%%%%%%%%%%%%%%%%%%%%%%%%%%%%%%%%%%%%%
\section{Summary and Comments}
\label{summary}
\setcounter{equation}{0}

The Racah and the $q$-Racah polynomials are the most generic members of
the orthogonal polynomials of a discrete variable satisfying second order
difference equations. By deforming the discrete quantum mechanical systems
governing these polynomials in terms of degree $\ell$ eigenpolynomials,
the exceptional Racah and $q$-Racah polynomials are obtained as the main
part of eigenfunctions of the deformed systems, which are shape invariant
and exactly solvable. By certain limiting procedures, the exceptional dual
($q$)-Hahn polynomials and the exceptional little $q$-Jacobi polynomials
are derived. The deformation process goes parallel with that for the
exceptional Wilson and Askey-Wilson polynomials.
Some of the characteristics of the quantum mechanics with real shifts
are the cause of complications which led to the delayed discovery.
The method of deriving the exceptional polynomials is new to the theory
of orthogonal polynomials.
As for the parameter ranges in which the orthogonality weight functions
are positive, we have made a quite conservative arguments. It is quite
possible that for a fixed $\ell$ the valid parameter range could be
enlarged than those given in the text.
On the other hand, the difference equations for the original and the
exceptional orthogonal polynomials, \eqref{forwardrel}--\eqref{difeqP},
\eqref{FlPln=flnPln}--\eqref{difeqPl} and
\eqref{FhatPn=Pln}--\eqref{BhatPln=Pn} are purely algebraic and they hold
for any parameter values.

With the understanding of all the generic exceptional orthogonal
polynomials as solutions of exactly solvable quantum mechanical systems,
the next challenge would be the construction of the exceptionals of
various reduced cases, for example, the Morse potential, the
Meixner-Pollaczek and the Krawtchouk cases, etc.
Finding multivariable generalisation is truly interesting but its
feasibility is as yet unclear.

%%%%%%%%%%%%%%%%%%%%%%%%%%%%%%%%%%%%%%%%%%%%%%%%%%%%%%%%%%%%%%%
%                                                             %
%  Acknowledgments                                            %
%                                                             %
%%%%%%%%%%%%%%%%%%%%%%%%%%%%%%%%%%%%%%%%%%%%%%%%%%%%%%%%%%%%%%%
\section*{Acknowledgements}

R.\,S. is supported in part by Grant-in-Aid for Scientific Research
from the Ministry of Education, Culture, Sports, Science and Technology
(MEXT), No.19540179.

%%%%%%%%%%%%%%%%%%%%%%%%%%%%%%%%%%%%%%%%%%%%%%%%%%%%%%%%%%%%%%%
%                                                             %
%  References                                                 %
%                                                             %
%%%%%%%%%%%%%%%%%%%%%%%%%%%%%%%%%%%%%%%%%%%%%%%%%%%%%%%%%%%%%%%

\end{document}